\documentclass[journal]{IEEEtran}
\usepackage[utf8]{inputenc}
\usepackage{cite}
\usepackage{setspace}
\usepackage{epsf}\epsfverbosetrue
\usepackage{graphics,epsfig}
\usepackage{graphicx,epsfig}
\usepackage{epsfig}
\usepackage{multirow}
\usepackage{alltt}
\usepackage{subfigure}

\usepackage{mathtools}

\usepackage{tcolorbox}
\usepackage{float}
\usepackage{url}
\usepackage{longtable}
\usepackage{graphicx}
\usepackage{verbatim} 
\usepackage{footnote} 
\usepackage{latexsym} 
\usepackage{sidecap} 
\usepackage{wrapfig}
\usepackage{eso-pic}
\usepackage{fix-cm}
\usepackage{algorithm}
\usepackage{algpseudocode}
\usepackage{color}
\usepackage{wrapfig}
\usepackage{multicol}
\usepackage{flowchart}
\usetikzlibrary{shapes,arrows,matrix,decorations.pathreplacing,shapes.geometric,positioning,calc}

\usepackage{layout}

\usepackage{amsmath}
\usepackage[
  locale = DE 
]{siunitx}

\usepackage{textcomp}
\usepackage{multirow}
\usepackage{mathtools}
\usepackage{soul} 
\usepackage{amsmath} 
\usepackage{verbatim}
\usepackage{csvsimple} 
\usepackage{array}
\usepackage{subfig}
\usepackage[normalem]{ulem}
\usepackage{booktabs}
\newcolumntype{P}[1]{>{\RaggedRight\arraybackslash}p{#1}}
\newcommand{\tabitem}{\textbullet~~}

\usepackage{pifont}
\newcommand{\tick}{\ding{52}}%
\newcommand{\cross}{\ding{55}}%
\newcommand{\mult}{\ding{93}}%

\tikzstyle{decision}=  [diamond, draw, fill=blue!50]
\tikzstyle{line}=      [draw,-stealth, line width= 0.4mm]
\tikzstyle{elli}=      [draw,ellipse,  fill=red!50,minimum height=5mm, text width=5em,  text centered]
\tikzstyle{block}=     [draw,rectangle,fill=red!45,  rounded corners,minimum height=13mm,text width=8em,  text centered]
\tikzstyle{medblock}=  [draw,rectangle,fill=blue!40, rounded corners,minimum height=18mm,text width=7em,  text centered]
\tikzstyle{childblock}=[draw,rectangle,fill=green!40,rounded corners,minimum height=25mm,text width=7em,  text centered]
\tikzstyle{MLblock}=   [draw,rectangle,fill=violet!40,rounded corners,minimum height=13mm,text width=3.3em,text centered]

\newcommand{\mubcom}[1]{\textcolor{black}{#1}}

\begin{document}

\title{Privacy Preserving Machine Learning for Electric Vehicles: A Survey}

\author{Abdul~Rahman~Sani,~Muneeb~Ul~Hassan,~Longxiang Gao,~and~Jinjun~Chen%
\IEEEcompsocitemizethanks{\IEEEcompsocthanksitem A.R. Sani and J. Chen are with the Swinburne University of Technology, Hawthorn VIC 3122, Australia (Email: abdulrahmansani.271@gmail.com; jinjun.chen@gmail.com)}
\IEEEcompsocitemizethanks{\IEEEcompsocthanksitem M. Ul Hassan is with Deakin University, Australia (Email: muneebmh1@gmail.com)}
\IEEEcompsocitemizethanks{\IEEEcompsocthanksitem L. Gao is with Qilu University of Technology, China and Shandong Computer Science Center, China (Email: gaolx@sdas.org)}}
\maketitle

\begin{abstract}
In the recent years, the interest of individual users in modern electric vehicles (EVs) has grown exponentially. An EV has two major components, which make it different from traditional vehicles, first is its environment friendly nature because of being electric, and second is the interconnection ability of these vehicles because of modern information and communication technologies (ICTs). Both of these features are playing a key role in the development of EVs, and both academia and industry personals are working towards development of modern protocols for EV networks. All these interactions, whether from energy perspective or from communication perspective, both are generating a tremendous amount of data every day. In order to get most out of this data collected from EVs, research works have highlighted the use of machine/deep learning techniques for various EV applications. This interaction is quite fruitful, but it also comes with a critical concern of privacy leakage during collection, storage, and training of vehicular data. Therefore,  alongside developing machine/deep learning techniques for EVs, it is also critical to ensure that they are resilient to private information leakage and attacks. In this paper, we begin with the discussion about essential background on EVs and privacy preservation techniques, followed by a brief overview of privacy preservation in EVs using machine learning techniques. Particularly, we also focus on an in-depth review of the integration of privacy techniques in EVs and highlighted different application scenarios in EVs. Alongside this, we provide a a very detailed survey of current works on privacy preserving machine/deep learning techniques used for modern EVs. Finally, we present the certain research issues, critical challenges, and future directions of research for researchers working in privacy preservation in EVs. 

\end{abstract}









\section{Introduction}
\begin{figure*}
  \includegraphics[scale=0.19]{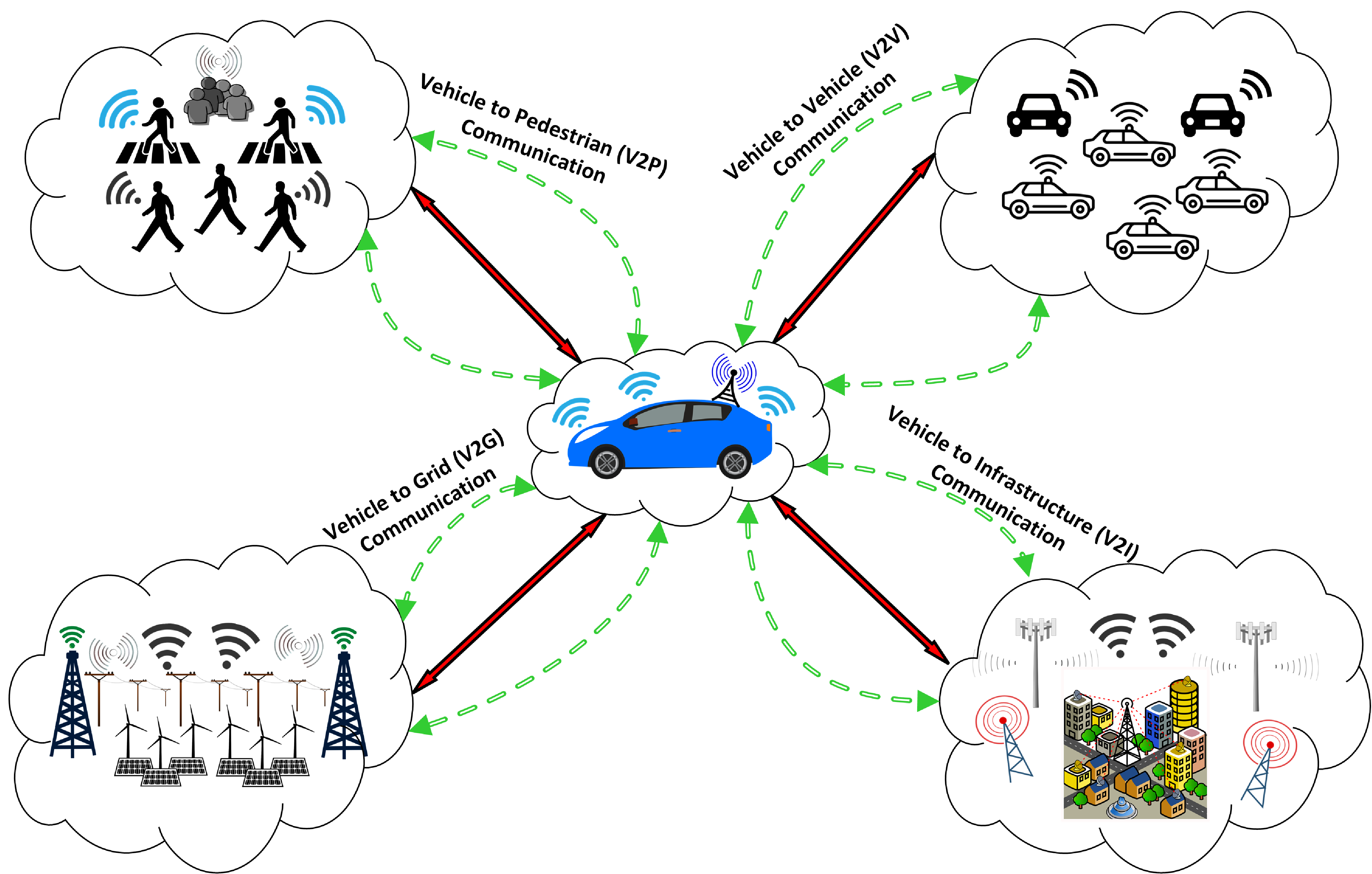}
  \centering
  \caption{\mubcom{Graphical Illustration of Vehicle to Everything (V2X) communication including the communication of Electric Vehicles with Pedestrians, Vehicles, Grid, and Infrastructure (adapted from~\cite{intref04}).}}
  \label{fig:v2x}
\end{figure*}

The advancements in intelligent transport systems (ITS)  alongside the growing need to eliminate the greenhouse gas emission caused by traditional transportation vehicles has led researchers and industrialists to consider the notion of modern electric vehicles (EVs) as an alternate transportation medium. Since modern EVs interact with each other via advanced communication networks in order to safety and critical information (e.g., road, environment, \& traffic assessment conditions), thus, they are being considered as a secure and trustworthy mode of transportation. According to a report by `International Energy Agency’, the total EVs around world will increase up to 230 million in number by the end of year 2030~\cite{intref01}. In order to provide this secure driving and travelling experience, it is important for each EV to connect with its surroundings and the controlling bodies. Modern EVs are connected with each other and also with their surroundings, which form a huge network comprises of multiple communication paradigms, such as vehicle-to-grid (V2G), vehicle-to-vehicle (V2V), vehicle-to-pedestrian (V2P), and vehicle-to-infrastructure (V2I) communication. All these paradigms are combined to form a notion for EVs named as vehicle-to-everything (V2X) communications~\cite{intref04}. A graphical illustration of V2X communication has been presented in Fig.~\ref{fig:v2x}.\\
V2X communication supports a diverse range of applications of EV, such as collision alerts, lane change alerts, traffic information and data sharing, navigation information etc~\cite{intref05}. Similarly, the data generated from these EVs will also be used for certain other useful purposes as well, such as navigation safety, identifying bottlenecks, optimising routes, and other tasks involving traffic management, accident prevention, etc. It has been estimated that an average EV produces around 30 Tera Bytes (TBs) of data each day~\cite{intref02}.  This huge amount of data is shared with the connected EVs, control servers, grid utilities, and other corresponding bodies so that timely actions are taken accordingly.\\
In order to get the most benefits out of this data, it is critically significant to utilize this data in an efficient manner. To ensure this efficient usage, machine learning models came as a rescuers, which enabled EV researchers to automate majority of tasks involved in EVs decision making~\cite{intref03}. Unlike the traditional fossil fuel based vehicles, modern EVs have two important characteristics: automation and cooperation~\cite{compsur14}. Major part of these characteristics is governed due to integration of machine /deep learning models with EV scenarios. Considering this, it is evident that machine learning will have a prime role towards development of an autonomous communication and decision network for future EVs. \\

\begin{table}[ht]
\begin{center}
\small
 \centering
  \captionsetup{labelsep=space}
 \captionsetup{justification=centering}
 \caption{\textsc{\\ \footnotesize{List of Acronyms Used in the Article.}}}
  \label{tab:acrtable}
  \begin{tabular}{|P{1.5cm}|P{6.3cm}|}
  	\hline
  	\rule{0pt}{2ex}
	\textbf{Acronyms} &\textbf{Definitions}\\
  	\hline
  	\rule{0pt}{2ex} 
  	 CNN & Convolutional neural networks\\
  	\hline
  	\rule{0pt}{2ex} 
  	 EV & Electric Vehicles\\
  	\hline
  	\rule{0pt}{2ex} 
  	 ITS & Intelligent transport systems\\
  	\hline
  	\rule{0pt}{2ex} 
  	 V2V & Vehicle-to-vehicle\\
  	\hline
  	\rule{0pt}{2ex} 
  	 V2I & Vehicle-to-infrastructure\\
  	\hline
  	\rule{0pt}{2ex} 
  	 V2X & Vehicle-to-everything\\
  	\hline
  	\rule{0pt}{2ex} 
  	 V2P & Vehicle-to-pedestrian\\
  	\hline
  	\rule{0pt}{2ex} 
  	 V2G & Vehicle-to-grid\\
  	\hline
  	\rule{0pt}{2ex} 
  	 TB & Tera Bytes\\
  	\hline
  	\rule{0pt}{2ex} 
  	 VANETS & Vehicular ad hoc networks\\
  	\hline
  	\rule{0pt}{2ex} 
  	 ICT & Information and communication technologies\\
  	\hline
  	\rule{0pt}{2ex} 
  	 IoT & Internet of Things\\
  	\hline
  	\rule{0pt}{2ex} 
  	 ICE & Internal combustion engine\\
  	\hline
  	\rule{0pt}{2ex} 
  	 OBU & On-board Unit\\
  	\hline
  	\rule{0pt}{2ex} 
  	 RSU & Road-side Unit\\
  	\hline
  	\rule{0pt}{2ex} 
  	 SVM & Support vector machine\\
  	\hline
  	\rule{0pt}{2ex} 
  	 ANN & Artificial neural networks\\
  	\hline
  	\rule{0pt}{2ex} 
  	 k-NN & k-Nearest neighbours\\
  	\hline
  	\rule{0pt}{2ex} 
  	 PCA & Principal component analysis\\
  	\hline
  	\rule{0pt}{2ex} 
  	 DQN & Deep Q-network\\
  	\hline
  	\rule{0pt}{2ex} 
  	 PII & Personally identifiable information\\
  	\hline
  	\rule{0pt}{2ex} 
  	 IDS & Intrusion detection systems\\
  	\hline
  	\rule{0pt}{2ex} 
  	 EDL & Energy demand learning\\
  	\hline
  	\rule{0pt}{2ex} 
  	 IoEV & Internet of Electric Vehicles\\
  	\hline
  	\rule{0pt}{2ex} 
  	 CIDS & Collaborative IDS\\
  	\hline 
  	\rule{0pt}{2ex} 
  	 FL & Federated Learning\\
  	\hline 
  	\rule{0pt}{2ex} 
  	 PoI & Point of interests\\
  	\hline 
  	\rule{0pt}{2ex} 
  	 LBS & Location based services\\
  	\hline 
  	\rule{0pt}{2ex} 
  	 IoV & Internet of Vehicles\\
  	\hline 
  	\rule{0pt}{2ex} 
  	 FCC & Federal Communication Commission\\
  	\hline 
  	\rule{0pt}{2ex} 
  	 DSRC & Dedicated short range communication\\
  	\hline 
  	\rule{0pt}{2ex} 
  	 FedAvg & Federated Averaging\\
  	\hline 
  	\rule{0pt}{2ex} 
  	 FedAG & FedAvg-Gaussian\\
  	\hline 
  	\rule{0pt}{2ex} 
  	 PPML & Privacy preserving machine learning\\
  	\hline 
  	\rule{0pt}{2ex} 
  	 SARSA & State–action–reward–state–action\\
  	\hline
  	\rule{0pt}{2ex} 
  	 TA & Trusted authority\\
  	\hline
  	\rule{0pt}{2ex} 
  	 ZEV & Zero-emission vehicles\\
  	\hline 	
  	\end{tabular}
  \end{center}
\end{table}


Despite of all these advantages, machine learning models are not 100\% secure and are vulnerable to various issues, in which privacy leakage issues is one of the most prominent one~\cite{compsur10}. It is because of the reason that machine learning techniques use such data which contains individually identifiable data (i.i.d) parameters of participants, which can lead to identification of a certain individual. Similarly, in case of machine learning based EV models, the data is not 100\% secure and can cause leakage of privacy if not handled properly. In order to eradicate such privacy issues, it is important to integrate some notion of privacy in traditional machine learning models. Researchers are actively working towards this integration, as some of the research works are working over integration of differential privacy preservation with machine learning to ensure that the model does not learn much about an individual. Similarly, some research works are developing decentralized notion of machine learning to avoid collection of data in a centralized database, etc. Thus, a quest to develop private and secure machine learning models for EVs is ongoing.  Thus, in this paper, we provide a comprehensive survey to show the importance of privacy in EV machine learning alongside providing a thorough analysis of present and prospective future works integrating the notion of privacy preserving machine learning (PPML) in EV scenarios.

\begin{table*}[htbp]
\begin{center}
 \centering
 \scriptsize
 \captionsetup{labelsep=space}
 \captionsetup{justification=centering}
 \caption{\textsc{\\Comparison and Summary of Previous Survey Articles on PPML, Privacy in EVs, Integration Scenarios of PPML in EVs Along With Major Contributions. \newline \tick ~shows that given topic is discussed, \cross ~shows that highlighted topic is not covered/discussed, and \mult ~shows that given topic is partially discussed.}}
  \label{tab:surveytable}
  \begin{tabular}{|P{2.1cm}|P{0.5cm}|P{0.5cm}|P{5.6cm}|P{1.1cm}|P{1.1cm}|P{1.1cm}|P{1.2cm}|P{1.3cm}|}
  	\hline
  	\rule{0pt}{2ex}
\centering \bfseries Application Scenario & \centering \bfseries Ref & \centering \bfseries Year & \centering \bfseries Major Contribution & \bfseries Discussed Privacy in EVs & \bfseries Discussed ML in EVs & \bfseries Discussed Private ML & \bfseries Integration Scenarios of PPML in EVs & \bfseries Existing Works for PPML in EVs \\
\hline

\multirow{4}{*}{\parbox{2cm}{\centering \textbf{V2G Networks in Smart Gird}}}

\rule{0pt}{2ex}
& \cite{compsur09} & 2016 & An investigation of privacy preservation problems and approaches in V2G networks & \tick & \cross & \cross & \cross & \cross \\
\cline{2-9}

& \cite{compsur17} & 2023 & A comprehensive survey for analysis of security challenges and countermeasures in V2G networks for smart grids. & \mult & \cross & \cross & \cross & \cross \\
\cline{2-9}
\hline

\rule{0pt}{2ex}
\centering \textbf{Vehicular and Mobile Communications} & \cite{compsur12} & 2018 & A comprehensive overview covering the aspect of privacy and security in various location-based services for mobile and vehicular networks & \tick & \cross & \cross & \cross & \cross \\
\hline

\multirow{4}{*}{\parbox{2cm}{\centering \textbf{Location Privacy in VANETs}}}

\rule{0pt}{2ex}
& \cite{compsur01} & 2019 & A thorough survey on deployment of location privacy models in in vehicular networks & \tick & \cross & \cross & \cross & \cross \\
\cline{2-9}

& \cite{compsur02} & 2019 & A thorough survey covering VANETs from the perspective of security, privacy and trust management & \tick & \cross & \cross & \cross & \cross \\
\cline{2-9}
\hline
\rule{0pt}{2ex}
\centering \textbf{Cyber Physical Systems} & \cite{compsur03} & 2020 & A comprehensive survey on application of differential privacy techniques in cyber physical systems & \tick & \cross & \cross & \cross & \cross \\
\hline

\rule{0pt}{2ex}
\centering \textbf{Privacy Attack in ML} & \cite{compsur13} & 2020 & In-depth analysis of more than technical works affiliated to privacy attacks in machine learning scenarios & \cross & \cross & \tick & \cross & \cross \\
\hline

\rule{0pt}{2ex}
\centering \textbf{6G Communication Networks} & \cite{compsur10} & 2020 & A comprehensive survey on privacy issues in machine learning from perspective of 6G networks and technologies & \mult & \mult & \tick & \cross & \cross \\
\hline

\multirow{10}{*}{\parbox{2cm}{\centering \textbf{IoT Networks}}}

\rule{0pt}{2ex}
& \cite{compsur06} & 2020 & An overview of IoT Networks security and privacy requirement on the basis of machine learning and certain deep learning methods & \cross & \cross & \tick & \cross & \cross \\
\cline{2-9}

\rule{0pt}{2ex}
& \cite{compsur07} & 2020 & A detailed study on resources management for IoT networks mechanisms, alongside discussing the use cases for machine \& deep learning in resources allocation & \cross & \cross & \mult & \cross & \cross \\
\cline{2-9}

\rule{0pt}{2ex}
& \cite{compsur08} & 2021 & A detailed study for privacy preserving IoT applications empowered by federated learning and taxonomy of FL over IoT Networks & \tick & \mult & \tick & \cross & \cross \\
\cline{2-9}

\rule{0pt}{2ex}
& \cite{compsur05} & 2021 & A survey on integration of federated learning in internet of things services and FL based IoT applications & \mult & \mult & \tick & \cross & \cross \\
\cline{2-9}

\hline

\multirow{4}{*}{\parbox{2cm}{\centering \textbf{Vehicular Networks}}}

\rule{0pt}{2ex}
& \cite{compsur14} & 2020 & A detailed overview of various adversarial attacks, taxonomy and defence methods for CAV machine learning & \mult & \cross & \cross & \cross & \cross \\
\cline{2-9}

& \cite{compsur15} & 2021 & A detailed survey of vehicular communication technologies, applications, vulnerabilities, and solutions based on machine learning and blockchain against cybersecurity attacks in vehicular networks & \mult & \tick & \cross & \cross & \cross \\
\cline{2-9}
\hline


\rule{0pt}{2ex}
\centering \textbf{IoV Communication} & \cite{compsur11} & 2021 & An overview of theoretical foundations of machine learning to solve security and privacy issues in IoV applications & \tick & \tick & \mult & \cross & \cross \\
\hline

\rule{0pt}{2ex}
\centering \textbf{Internet of Vehicles} & \cite{compsur16} & 2023 & They proposed integrating machine learning with blockchain to create a secure and scalable data exchange platform for the Internet of Vehicles. & \cross & \mult & \cross & \cross & \cross \\
\hline

\rule{0pt}{2ex}
\centering \textbf{Wireless Communication Networks} & \cite{compsur04} & 2021 & A thorough review work on various distributed machine learning techniques from the perspective of wireless networks and its applications & \cross & \mult & \tick & \cross & \cross \\
\hline

\rule{0pt}{2ex}
\centering \textbf{UAVs in 5G and Beyond Networks} & \cite{compsur18} & 2024 & A comprehensive review on applications, challenges, and future directions of UAV integration within 5G and beyond networks, especially for the Internet of Things. & \mult & \cross & \cross & \cross & \cross \\
\hline

\rule{0pt}{2ex}
\centering \textbf{Vehicle as a service in smart cities} & \cite{compsur19} & 2024 & An overview of Vehicles-as-a-Service as a new paradigm for smart city infrastructure, leveraging vehicles capabilities for distributed data processing and reducing infrastructure costs. & \mult & \cross & \cross & \cross & \cross \\
\hline

\rule{0pt}{2ex}
\centering \textbf{Privacy Preserving Electric Vehicles} & This Work & 2021 & An in-depth survey integration of privacy preserving machine learning in electric vehicles from the viewpoint of integration scenarios, technical works, and prospective challenges and directions. & \tick & \tick & \tick & \tick & \tick \\
\hline
 \end{tabular}
  \end{center}
\end{table*}

\subsection{Comparison with Related Survey Articles}

Our current survey work on PPML for EVs is significantly different from all past research works, as we thoroughly discuss the area of PPML from perspective of integration scenarios, existing technical works, and a thorough analysis of prospective challenges and future directions.  From the perspective of previous literature, an extensive number of survey works has been presented targeting different scenarios of EVs (see Table~\ref{tab:surveytable}). For example, one of the pioneering survey covering the domain of preservation of privacy in V2G networks have been presented by Han and Xiao in~\cite{compsur09}. Authors first presented a detailed framework for V2G communications and then discussed the privacy attacks which can cause data leakage in this framework alongside discussing goals and approaches that helped tackle these issues. Similarly, Rajasekaran~\textit{et al.} in \cite{compsur17} examined security challenges in V2G networks, analyzing vulnerabilities and potential countermeasures for confidentiality, integrity, and other security objectives.  The authors explored various security threats and recommended methods to safeguard communication between EVs and the smart grid.

Another interesting work targeting the domain of privacy and security in location based services from the perspective of mobile and vehicular communication has been presented by Asuquo~\textit{et al.} in~\cite{compsur12}. Authors formulated the article on the advantages of location based services and then highlighted the concerns in these services from security and privacy perspective. A very similar work covering the aspect of deployment of location privacy models in vehicular ad hoc networks (VANETs) have been presented by Talat~\textit{et al.} in~\cite{compsur01}. Authors extensively classified the threat models in EVs alongside covering the domain of privacy preservation strategies. Another similar work focusing over the advances in VANETS from security, privacy, and trust management perspective have been discussed by authors in~\cite{compsur02}. The technical article summarizes a thorough discussion over these issues starting with the preliminaries of VANETs and then providing an expert opinion that how the issues can be tackled. \\
From the viewpoint of privacy preservation via differential privacy, a very detailed work has been published by authors in~\cite{compsur03}. In the article, authors work over demonstration of the use of differential privacy perturbation strategy in cyber physical systems including intelligent transportation systems (ITS). From the viewpoint of privacy attack in computing and machine learning networks, a very detailed survey has been presented by Rigaki and Garcia~\cite{compsur13}. Authors provide a thorough analysis that how machine learning models are prone to privacy attacks and what are prospective ways to mitigate such risks. Another similar covering the integration of privacy and machine learning from perspective of 6G networks have been presented by Sun~\textit{et al.} in~\cite{compsur10}. The article first provides detailed analysis of privacy attacks on machine learning models and afterwards provide discussion about protection strategies that can be employed. \\
From the perspective of Internet of Things (IoT) networks, machine learning, and privacy, a few research works~\cite{compsur06, compsur07, compsur08, compsur05} have been published in the recent past. The first work covering the use of machine learning in the security and privacy of IoT networks have been presented by Hussain~\textit{et al.} in~\cite{compsur06}. Another work focusing over machine learning based resource management in IoT and cellular networks have been carried out by authors in~\cite{compsur07}. A very detailed work showing the use cases and application of federated learning in IoT networks has been published by Khan~\textit{et al.} in~\cite{compsur08}. Similarly, another significant work covering the similar domain of federated learning and services and applications of IoT have been presented by authors in~\cite{compsur05}. Alongside this, a thorough work discussing the overview of adversarial attacks alongside defence mechanisms in autonomous and interconnected vehicles have been presented by Qayyum~\textit{et al.} in~\cite{compsur14}. A distinct work leveraging the use and integration of blockchain and machine learning in modern vehicular networks have been presented by Dibaei~\textit{et al.} in~\cite{compsur15}. The work investigated the future prospects and challenges of this integration in a detailed manner. Similarly, a very interesting work discussing the theoretical foundation of security and privacy of machine learning communication of vehicular networks have been presented by authors in~\cite{compsur11}. In order to discuss the potential of distributed machine learning in modern vehicular networks, a very comprehensive survey has been Hu~\textit{et al.} in~\cite{compsur04}. \\

Zamanirafe ~\textit{et al.} in~\cite{compsur16} explored how combining machine learning (ML) with blockchain can address security and scalability challenges in the Internet of Vehicles (IoV). The authors proposed a system where vehicles train ML models locally and collaborate securely via blockchain to create a more robust and trustworthy IoV ecosystem.

Banafa ~\textit{et al.} in~\cite{compsur18} explored integrating UAVs into next-generation wireless networks (5G and beyond). They examined applications in IoT, highlighting challenges like privacy and energy limitations, and proposed promising research directions for reliable UAV communication.

In another study authors proposes VaaS, a novel approach leveraging vehicles sensing, communication, and computing capabilities (SCCSI) to create a distributed service network for smart cities. This SCCSI-powered VaaS architecture reduces infrastructure costs and improves service delivery with lower latency, but necessitates further research on security, privacy, and incentive mechanisms ~\cite{compsur19}.

\textit{However, to best of our knowledge, there is no research survey article present in the literature till now that extensively discuss the PPML and its implementation scenarios from the perspective of modern EVs.}

\subsection{Contributions of Our Survey Article on PPML in EVs}
Since a certain number of surveys articles which are published in the past have discussed some certain aspects of privacy preservation in electric vehicles. However, to our best knowledge, no such survey that provides information about integration of private machine learning in EVs has been published so far. In this paper, we provided a state-of-art survey over current works on PPML for EVs from various perspectives. In summary, the major contributions of our work are mentioned as follows:

\begin{itemize}
\item We provide a detailed review of previous survey articles on privacy preservation in EVs and communication networks.
\item We provide a thorough discussion about the integration of PPML in various applications scenarios of EVs.
\item We survey the works done over implementation of PPML in the domain of EVs.
\item We outline certain challenges, open issues, and prospective reseach directions for future in PPML for EVs.
\end{itemize}

\subsection{Article Structure}
A list of acronyms that we have used in this survey article has been given in Table~\ref{tab:acrtable}. The remaining part of the paper is organized in different sections as follows: In section II we provide an overview of electric vehicles, VANETs, machine learning and privacy preservation techniques. Motivation behind PPML for EVs and advantages of privacy integration in EVs is described in section III. While in the section IV, we provide a detailed survey of privacy integration scenarios in electric vehicles. Moreover, the current works on PPML for EVs have been covered in section V. In section VI, we highlight various research issues, important challenges, and future directions for research. Finally, we provide conclusion our survey article in section VII.


\section{Preliminaries of Privacy and Machine Learning in EVs}

The interest in EVs is growing rapidly, which also raises concerns regarding leakage of privacy in modern EVs. In this section, we first discuss the preliminaries and basics of the article such as functioning of EVs, vehicular ad-hoc networks (VANETs), machine learning, and privacy preservation. Afterwards, we discuss the need of privacy preservation in EVs alongside discussing certain PPML strategies. 
\subsection{Electric Vehicles}
The concept of EVs is not completely novel, as it has been in discussion among the community for a long time. Even, the history reports that the development of first EV is dated back to 1984~\cite{prelev02}. Afterwards, in the 19th century,  a large number of companies worked over the development of EVs, especially in France, America, and Britain. Although, due to limitation of batteries and communication technologies alongside the rapid developments in internal combustion engine (ICE) lead to the downfall of EVs around the year 1930. However, since the beginning of 21st century, the interest of scientific community towards development of EVs increase again because of two major reasons: Firstly, the attention of scientific community towards development of a zero-emission vehicles (ZEVs) increased. Secondly, with the development of modern information and communication technologies (ICTs), everything is becoming interconnected, which lead towards formation of interconnected EVs with the help of V2X communication. Another critical factor that caused an increase in interest towards modern EVs, is that due to ICTs, a large number of EVs are capable of providing partial or complete autonomous experience for driving and other tasks~\cite{prelev01}. 
Apart from these capabilities, another direction, which attracted the attention of research and industry is the capability of EVs to store and share energy in a bidirectional manner at the time of need~\cite{prelev03}. This led to the development of a new field of grid-connected electric vehicles, in which EVs energy trading patterns are planned in a way that they help in enhancement of demand response of a specific area by maintaining the demand-supply curve of that specific region. All these reasons played their part towards the increased interest in modern EVs, and because of this interest,  it has been estimated that the number of EVs around the globe will increase up to 230 million by the end of 2030~\cite{intref01}.

\subsection{Vehicular Ad Hoc Networks (VANETs)}

As the number of interconnected EVs equipped with ICTs increases, the communication among them escalated to an extent, which lead towards development of a new type of networks named as VANETs~\cite{prelev04}. VANETs empower autonomy of EVs by aiding a large number of applications, such as real-time traffic monitoring, collision prevention, dynamic route scheduling, safety, security, etc. Typical VANETs used to have two major type of communications, named as V2V and V2I communication. In which vehicles are usually only connected with other vehicles and corresponding infrastructure, respectively. However, nowadays, VANETs are much more capable and provide the advantages of communication of EVs with other involved participants, such as pedestrian, smart grid, via V2P and V2G communication respectively. Combinedly, all these sort of communications are known as V2X communication, which describes the interconnection of vehicle with all other connected devices within a single umbrella. A detailed discussion of VANETS will be out of scope of this article, however, in order to provide our readers a brief overview, we discuss three prominent parts of VANETs named as  on board unit (OBU), road side unit (RSU), and trusted authority (TA).
\subsubsection{On-Board Unit (OBU)}
The unit which is mounted over the EV is known as OBU, this unit acts as controller, which is used to carry out various tasks such as gathering, processing, transmission, and management of data generated or received at the vehicular end~\cite{prelev05}. This also acts as a major communication devices, which is responsible to receive and transmit data between surrounding vehicles’ OBUs and RSUs. Multiple sensors are usually mounted over OBU to perform certain critical operations, such as (a) measuring and reporting the important parameters of an EV, such as health of engine and breaks, alongside light and surrounding conditions. (b) regularly monitoring the detecting the parameters related to collision avoidance (c) tracking and reporting the critical parameters of EV, such as location, GPS orientation, gyros, etc. (d) managing the infotainment, multimedia, and weather reporting system. Alongside providing all these services to EV, the OBU is responsible to receive the real-time updated from neighbouring EVs, RSUs, and pedestrians in order to take timely actions where required. 
\subsubsection{Road Side Unit (RSU)}
The RSU, which is usually connected with surrounding EVs and RSUs is placed at certain road-side points as a static unit. RSU is equipped with transmitters and receivers that receive and transmit data to and from OBUs and Tas~\cite{prelev06}. Thus, basically serves as an intermediary communication medium between EVs’ OBUs and various controlling TAs. RSUs are usually placed at crucial sights at prescribed spacing in order to provide dependable network coverage and operational efficiencies. As a result, the distance value among neighbouring RSUs is maintained in a way that they still remain within the range of communication of other consecutive RSUs. RSUs is often used for certain other tasks, such as expanding the range of communication in the case of mobile vehicular networks, or to operate critical reporting programmes such as reporting of an accident or critical weather warning or forecasting, or just to give internet access to the neighbouring OBUs.

\subsubsection{Trusted Authority (TA)}
TA serves as the backbone of the whole ITS and is often linked to RSUs by optical fibre cables or wireless medium~\cite{prelev07}. TA is in charge of managing security and trusted communication in VANETs. Through RSUs, TA authenticates all network components within the specified area range. It is also in charge of identifying whether any OBU that is sending adversarial packets or subsequently cancelling the subject node during communication. TA is often located in the heart of a city and is controlled by state officials or any other regulatory body which is in-charge of controlling of EVs. Because a TA deals with vast amounts of data and computationally expensive cryptographic processes, it is necessary to have strong processing computational power and ample storing capacity to carry out data aggregation, data storage, and systematic analysis over data. Within a region, many TAs can be established to improve coverage, distribute control, and prevent a single point of failure. High-speed data lines for sharing real-time information amongst dispersed TA designs are usually deployed in order to facilitate the timely communication and decisions. 

\begin{figure*}
  \includegraphics[scale=0.5]{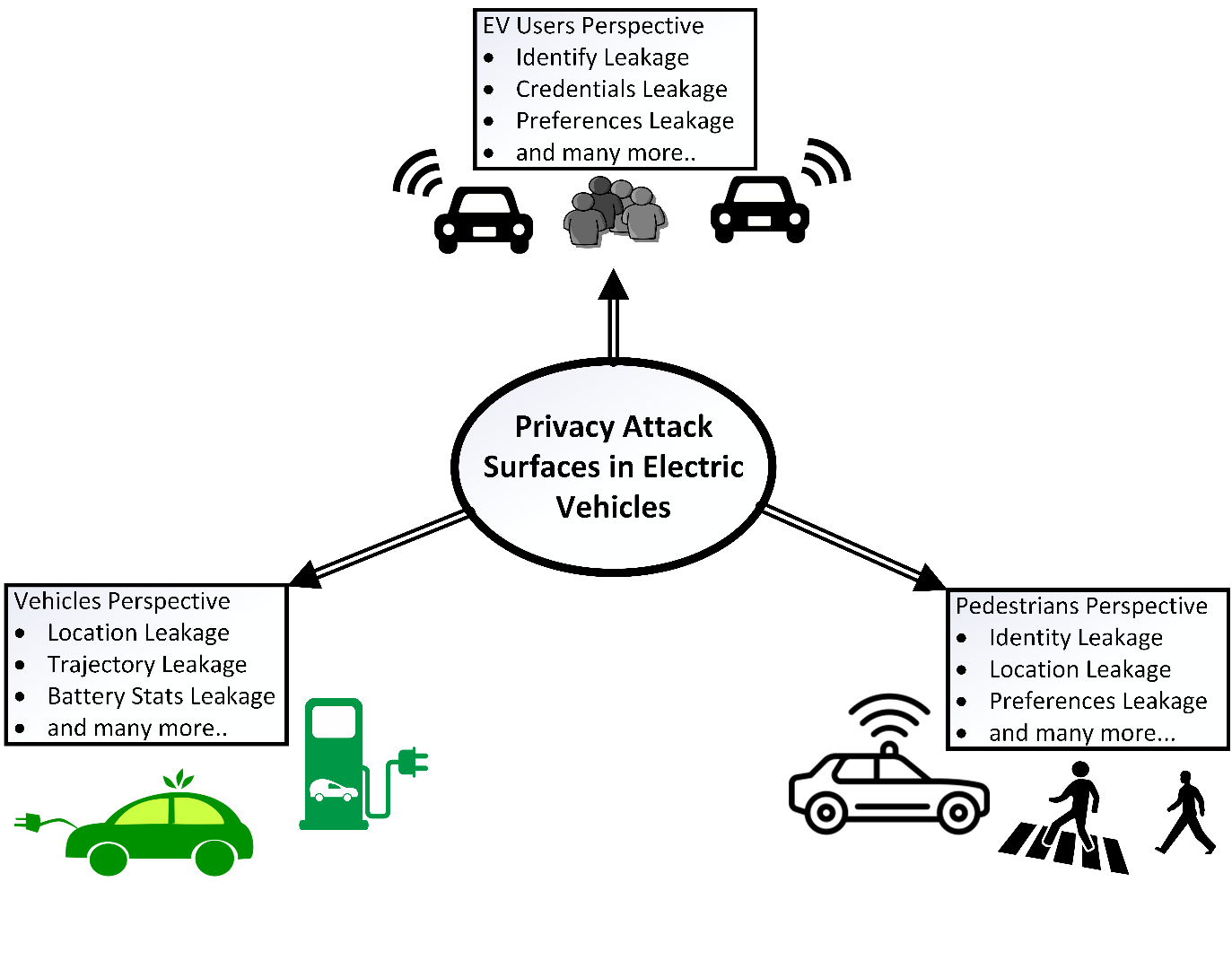}
  \centering
  \caption{\mubcom{Classification of Privacy Attack Surfaces in Electric Vehicles from perspective of vehicles, EV users, and pedestrians.}}
  \label{fig:attackEV}
\end{figure*}

\subsection{Machine Learning}
Machine learning algorithms received a huge attention from research and industrial community because of their applications in almost every field of life ranging form healthcare to modern EVs. That is because of the reason that machine learning algorithms provides computer with an artificial intelligence, which makes the device capable of learning from environment and taking autonomous decisions. A number of research works has been published in this field and a vast number of sub-field and sub-algorithms have been derived out of conventional machine learning algorithms, such as reinforcement learning, deep learning, etc. A detailed discussion about all of these algorithms will be out of scope of our article. However, to provide our readers a basic perception about the functioning and categorization these models, we provide a quick overview of four types of models of machine learning, which generally cover the majority of models of machine learning which are useful in understanding the perspective of machine learning in EVs. 
\subsubsection{Supervised Machine Learning }
One of the most prominent category in machine learning models is supervised machine learning, which basically deals with integration of ability of the model to learn by providing the model with the data alongside the desired outcomes or labels~\cite{prelev07}. This direction of training machine learning models by providing solutions and base data  is not new and has been in discussion since many decades~\cite{prelev08}. However, one cannot deny the importance of such models, as a well-trained supervised learning model can categorise the new output in a very efficient manner. The actual goal of supervised learning models is to develop and train such a function which maps input with the given outputs efficiently. In this way, an efficiently trained model will be able to predict the novel input with a precise and accurate manner. From perspective of EVs, such models are generally used in decision making processes, where an efficient and accurate decision is required. Moreover, a vast number of algorithms/models have been developed for supervised machine learning, however, some of the prominent ones are support vector machine (SVM), artificial neural networks (ANN), linear regression, k-nearest neighbours (kNN),  etc. 

\subsubsection{Unsupervised Machine Learning}
As suggested by the name, unsupervised machine learning is completely opposite to supervised machine learning from the perspective of provided data. In unsupervised machine learning, the model learns itself from the provided data by finding hidden patterns in it~\cite{prelev09}. Such type of models play an important role, when the aim is to group certain individuals or identify some anomalies. From perspective of EVs, they are usually used in anomaly detection, clustering, and visualization of the recoded data. Similar to supervised machine learning, a larger number of models have been designed for unsupervised machine learning as well ranging from k-means clustering to principality component analysis (PCA), etc.
\subsubsection{Reinforcement Learning}
This is a specialized type of machine learning which basically works over the phenomenon of trial, error, and reward~\cite{prelev10}. For instance, in reinforcement learning, a model is introduced to the desired dataset, from which it learns its attributes, and on the basis of these attributes, a trial is taken, and in case if the outcome is correct, then the model is rewarded. Otherwise, if the outcome/prediction is not right, then a punishment is given to the modelling agent. In this way, the model trains itself to map and incline towards the positive desired outcome alongside deviating from the undesired negative outcome. Some works also classify reinforcement learning as a sub-category of unsupervised learning~\cite{prelev11, prelev12, prelev13}. However, due to the nature of its application in EVs perspective, we discuss this domain as a distinct category. From the EVs perspective, these sort of models are usually used in optimization of consumer cost and other similar energy management operations. Some of the common reinforcement learning models can be named as Q-learning, State–action–reward–state–action (SARSA), deep Q-network (DQN), etc.

\subsection{Privacy Preservation Techniques}
Preserving one’s private information is a right of every individual, and sometime observing and reading data from various sources can leak private information. In order to overcome this issue, researchers have worked and are continuously working over development of efficient privacy preservation models. A brief discussion about all privacy protection techniques will be out of scope of this article. However, in this section, we discuss few major types which are being used by researchers in modern EV scenarios. 
\subsubsection{Differential Privacy}
The phenomenon of Differential privacy was introduced first by Cynthia Dwork in the year 2006 as a notion to protect privacy of data in a statistical databases scenario during query evaluation~\cite{prelev17}. Although, with time, differential privacy proved its effectiveness in a large number of other scenarios as well ranging from real-time reporting to geo-location protection. Similarly, in case of PPML in EVs, differential privacy has also proved its effectiveness and a plethora of research works have been published in this regard, which either used differential privacy in real-time EV reporting, or used it during learning data from EVs databases. 
\subsubsection{Homomorphic Encryption}
Since the advent of homomorphic encryption, a new direction of carrying out computation over the encrypted data has been established. Homomorphic encryption allows personals from academia and industry to carry out required computational over any data without finding/revealing the actual contents of that data~\cite{prelev15}. A vast number of works have been carried out from the context of integration of homomorphic encryption in real-life scenarios, such as location privacy protections, etc~\cite{prelev16}. Similarly, from the perspective of machine learning in EVs, homomorphic encryption can play a critical role in preserving privacy of EVs. This privacy preserving notion can protect privacy in multiple scenarios, such as real-time data transmission, collaborative learning, database analysis, etc.

\subsubsection{Decentralized Machine Learning}
Apart from the traditional privacy preserving notions, research works have emphasized that if one do not collect the private data from individuals and only carry out model training at local end devices in a decentralized manner, then privacy can be protected without integrating a specific privacy preserving notion. No doubt it is pretty effective to a specific extent as no data is being stored at a centralized server. However, it still is an under discussion topic in the scientific community because researchers are also integrating various privacy preservation strategies with decentralized learning to provide more privacy. Although from the perspective of decentralized learning, the most prominent one is federated learning, which was in 2016 by Google AI researchers as a notion to learn from their users in a decentralized manner~\cite{prelev18}. Alongside improving the communication, latency, and data storage cost, federated learning also provides privacy preserving learning as the centralized authority only send the model to learn from end users in a decentralized manner. Similarly, in case of EVs, a larger number of research works have been presented, which we discuss later in this article. 

\begin{table*}[ht]
\begin{center}
 \centering
 \footnotesize
 \captionsetup{labelsep=space}
 \captionsetup{justification=centering}
\caption{\textsc{\\ {Summary of Works integrating Privacy Preservation with Machine Learning Models}}}
  \label{tab:ppmltable}
 {\color{black} \begin{tabular}{|p{3cm}|p{0.6cm}|p{0.7cm}|p{2cm}|p{9cm}|}
  	\hline
\rule{0pt}{2ex}
\centering \bfseries Machine Learning Model & \centering \bfseries Ref. \# & \centering \bfseries Year & \centering \bfseries Privacy Technique & \bfseries Contribution of the Article \\
\hline
\rule{0pt}{2ex}
\centering \textbf{Federated Learning} & \cite{ppml01} & 2020 & $\varepsilon$-Differential Privacy & Integrated Local DP with federated learning (FedSGD) to enable the crowd-sourcing applications to train machine learning model by avoiding privacy discloser and reducing the communication overhead. \\
\hline
\rule{0pt}{2ex}
\centering \textbf{Deep Neural Networks} & \cite{ppml02} & 2019 & $\varepsilon$-Differential Privacy & Used layer-wise perturbation and DP mechanism to enhance the accuracy gap between privacy/non-privacy preserving deep neural networks. \\
\hline
\rule{0pt}{2ex}
\centering \textbf{K-Means Clustering} & \cite{ppml03} & 2016 & $\varepsilon$-Differential Privacy & Improved the interactive and non-interactive differentially private k-means clustering approaches based on the systemised error analysis.  \\
\hline
\rule{0pt}{2ex}
\centering \textbf{Linear Regression} & \cite{ppml04} & 2020 & Homomorphic Encryption & Integrated homomorphic encryption and data masking techniques to proposed a new privacy preserving linear regression protocol. \\
\hline
\rule{0pt}{2ex}
\centering \textbf{Support Vector Machine} & \cite{ppml05} & 2019 & Homomorphic Cryptosystem & Employed blockchain with homomorphic cryptosystem in order to build a safe, secure, and privacy preserving SVM training model for data sharing in IoT. \\
\hline
\rule{0pt}{2ex}
\centering \textbf{Artificial Neural Networks} & \cite{ppml06} & 2020 & $\varepsilon$-Differential Privacy & Differentially private ANN to accomplish the privacy of individual COVID-19 data by keeping utility closer to the baseline ANN models. \\
\hline
\rule{0pt}{2ex}
\centering \textbf{Logistic Regression} & \cite{ppml07} & 2019 & Homomorphic Encryption \& $\varepsilon$-DP & A novel scheme to support distributed logistic regression learning model by taking advantage of combination of Laplace perturbation and homomorphic encryption. \\
\hline
\rule{0pt}{2ex}
\centering \textbf{Naive Bayes} & \cite{ppml08} & 2018 & $\varepsilon$-Differential Privacy & A privacy preserving naive bayes classification scheme by utilizing DP techniques for data subsidized from multiple data sources. \\
\hline
\rule{0pt}{2ex}
\centering \textbf{Ensemble Classifier} & \cite{ppml09} & 2019 & Additive Secret Sharing & A privacy-preserving ensemble classification framework using additive secret sharing and edge computing techniques to protect individual face features and service provider privacy in face recognition. \\
\hline
\rule{0pt}{2ex}
\centering \textbf{Conventional Neural Networks} & \cite{ppml10} & 2019 & ($\varepsilon$,$\delta$)-Differential Privacy & Integrated differential privacy in CNN model by utilizing adaptive gradient descent with novel privacy budget allocation scheme. \\
\hline
\rule{0pt}{2ex}
\centering \textbf{Decision Tree} & \cite{ppml11} & 2020 & $\varepsilon$-Differential Privacy & A privacy-preserving decision tree model with high utility by leveraging differential privacy techniques with a new privacy budget allocation scheme. \\
\hline
\rule{0pt}{2ex}
\centering \textbf{Random Forest} & \cite{ppml12} & 2015 & $\varepsilon$-Differential Privacy & A new random forest model under the framework of DP  for privacy preserving data mining with high utility. \\
\hline
\rule{0pt}{2ex}
\centering \textbf{K-Nearest Neighbors} & \cite{ppml13} & 2017 & $\varepsilon$-Differential Privacy & Developed models for differentially private instance-based radius neighbors classifier (r-N) and k-NN with improved classification accuracy. \\
\hline
\rule{0pt}{2ex}
\centering \textbf{Reinforcement Learning} & \cite{ppml14} & 2021 & $\varepsilon$-Differential Privacy & An optimized privacy-preserving scheme leveraging DP and RL to acquire an optimum privacy budget distribution strategy for VANETs. \\
\hline
\rule{0pt}{2ex}
\centering \textbf{Deep Reinforcement Learning} & \cite{ppml15} & 2020 & $\varepsilon$-Differential Privacy & Developed a DRL based user profile perturbation methodology and integrated DP with DRL to protect user privacy against inference attacks for recommendation systems. \\
\hline
 \end{tabular}}
  \end{center}
\end{table*}

\section{Motivation and Requirement of Privacy Preserving Machine Learning in EVs}
In this section, we first discuss the basics of private preserving machine learning alongside discussion various research works which integrated privacy in machine learning. Afterwards, we highlighted the motivation and requirement of privacy in modern EVs. 

\subsection{Privacy Preserving Machine Learning}
With the rapid increase of data creation around the globe through internet, web, end users, etc., a huge amount of data is now available for data extraction techniques to learn users’ behaviour. Machine learning and its sub-types comes up as a first contender in such data extraction techniques, which allow computers to learn the attributes of users’ data and then perform tasks in an artificially intelligent manner~\cite{prelev20}. In machine learning, a number of steps from data collection to making predictions is involved, which is being carried out either at user end or at server. For instance, first of all, data is collected and prepared for training, which is also known as pre-processing. Afterwards, the corresponding model is chosen for the training process, which highly depends upon the type and requirement of application/outcome. After that, the model is trained, evaluated, and tuned according to the need. Finally, the trained model is given a new dataset to carry out prediction. All these steps involve a number of sub-steps and each of the sub-step has its corresponding processed involved. However, the common goal of all these steps is to carry out efficient predictions. \\
Similarly, if one applies a machine learning based solution to carry out prediction over any specific data, then one of the most critical thing that need to be taken care of it the data and user privacy~\cite{prelev21}. It is because the datasets which are usually used in machine/deep learning models contain a large amount of private and sensitive information that can cause leakage of an individuals’ privacy~\cite{prelev19}. For instance, a dataset collected from EVs in a specific area can have information about the visited locations, usernames, profiles, surnames, schedules, passwords, etc. This information can further be linked with other datasets to figure out medical histories, religious preferences, locational preferences, etc. In certain cases, this data can also be sold out to advertising companies to run personalized and targeted advertisements. Thus, one needs to protect the privacy of involved participants at the time of learning. \\
In order to do so, a plethora of works have been carried out by researchers, in which researchers worked over integration of various privacy preservation techniques with machine learning models. A detailed discussion of each such direction will be out of scope of this article. However, in order to give a brief overview, we highlight certain works in this section. A table summarizing the integration of privacy preservation with machine learning models in different works have been presented in Table~\ref{tab:ppmltable}. One of the significant work in this category has been carried out by Zhao~\textit{et al.} in~\cite{ppml01}, who integrated the notion of differential privacy with decentralized federated learning. The next interesting work integrating differential privacy in layer-wise deep neural networks has been presented in~\cite{ppml02}. From the perspective of integration of differential privacy with traditional k-means clustering, a comprehensive technical work has been presented by Su~\textit{et al.} in~\cite{ppml03}. Similarly, a work presenting private linear regression with the use of homomorphic encryption has been published by Qiu~\textit{et al.} in~\cite{ppml04}. Another work using homomorphic encryption with support vector machine (SVM) has been presented in~\cite{ppml05}. \\
Moving further towards another category of neural networks named as artificial neural networks (ANN), a comprehensive work integrating differential privacy with ANN has been presented by Sinha~\textit{et al.} in~\cite{ppml06}. An interesting work integrating Laplace perturbation and homomorphic encryption to preserve privacy in logistic regression based learning has been presented in~\cite{ppml07}. From the family of probabilistic classifiers, a work targeting integration of differential  privacy with naïve bayes classifier has been presented in~\cite{ppml08}. Alongside this, a work covering the notion of privacy preserving ensemble classification has been presented by Ma~\textit{et al.} in~\cite{ppml09}. The adaptation of a novel privacy budget allocation models in conventional  neural networks (CNN) has been carried out by researchers in~\cite{ppml10}. Similarly, another work focusing over the use of efficient differential privacy budget allocation for decision tree based learning has been presented by authors in~\cite{ppml11}. While tackling the issue of mining utility via random forest alongside leveraging the potential of differential privacy has been explored in~\cite{ppml12}. Another similar work integrating differential privacy in traditional k-nearest neighbour learning has been presented in~\cite{ppml13}. A novel work covering the notion of optimum budget allocation in VANETs for differentially private reinforcement learning has been published by authors in~\cite{ppml14}. Finally, a research article covering the notion of differentially private deep reinforcement learning has been presented by Xiao~\textit{et al.} in~\cite{ppml15}.\\
Overall, large number of works targeting the integration of privacy preservation techniques in machine/deep learning context has been carried out so far. Similarly, it will not be wrong to say that almost all major machine learning models have been integrated with modern notions of privacy in order to provide a private and privacy preserving data analysis. Therefore, in future real-time scenarios (such as EVs), it is becoming a need to use private machine learning models instead of traditional machine learning in order to enhance trust in the network.

\subsection{Privacy Attacks Surfaces in Electric Vehicles}
Due to the tremendous benefits of EVs, a large number of industries are moving towards development of modern EVs~\cite{prelev24}. As a result, attackers and adversaries have also become active and are trying to get unfair advantage of every possible loophole in the network. Thus, efforts need to be made to protect the privacy of EVs at every possible level. In order to give a comprehensive overview of privacy requirements, we divide them into three major perspectives, which covers almost all sort of privacy attacks and their defence measures. A graphical illustration of privacy attacks in EVs has been presented in~\ref{fig:attackEV}
\subsubsection{EV Users Perspective}
Modern EVs have improved connectivity due to the advances in communication technologies, such as high data rates, low latency, etc. This connectivity is aiding in accomplishment of a large number of tasks of EVs, such as enhancing driving experience by learning about surroundings, etc. Contrarily, this connectivity can also result in loss of privacy of EV users. For instance, the pervasively connected network creates novel possibilities for attackers and hackers to target and compromise the connected EVs. Attackers can target various misconfigurations, design bugs, implementations flaws, etc to infer into private information of EV users, which can lead to privacy leakage. E.g., if an EV is publicly sharing its presence to surrounding EVs, then an attacker can try to compromise the communication link or can take advantage of any bug in the network to figure out the owner of that particular EV~\cite{prelev23}. In this way, the attacker/hacker can easily identify the activities of the owner of that particular EV. For example, if the moving pattern of an EV gets compromised, one can infer that when the owner goes to work, comes homes, goes to gym, etc. This information can further be used to a large number of malicious purposes, such as theft, burglaries, etc. Even certain non-malicious companies could be interested in getting information about the movement patterns of EV owners, so that they can do targeted advertisement. E.g., if an owner is going to gym regularly, they can show fitness related advertisements to them. This can be broken down into many perspectives and privacy of EV users can be comprised at different levels. Therefore, the need to protect privacy of EVs users in a much efficient manner is required. 

\subsubsection{Vehicles Perspective}
From the perspective of vehicles, various types of vehicles can be a target to certain privacy attacks. Starting from autonomous vehicles, which are capable of driving themselves without involvement of any driver. One very famous application of an autonomous vehicle is `driverless taxis’, which is being operated in a number of countries nowadays~\cite{prelev22}. These autonomous EVs are continuously gathering information about their surroundings in order take efficient decisions. However, in the quest of taking autonomous decisions, these EVs are gathering way too private information about their surroundings. E.g., an EV usually scans its neighbouring vehicles with cameras mounted over them. This data is then sent to control centres as well to carry out future training of models. But on the other side, this data contains a large number of personal identifiers, which should not be disclosed to training bodies/centres. For instance, from the captured image of surrounding cameras from EVs, one can easily identify that a specific vehicle was present on the road at a particular instance of time. Similarly, one can identify the owner of that vehicle as well, and in this way one can easily investigate that a particular person went to a particular place at a specific time. Similarly, this can also be taken in a perspective that at the time of a specific event, which vehicles were present, which can lead to an investigation that the particular drivers of these vehicles saw the incident. This can lead to leakage of identity privacy and location privacy for the surrounding vehicles. Therefore, it is important for EVs to take appropriate measures before transmitting surrounding vehicular data.  

\subsubsection{Pedestrians Perspective}
As discussed earlier that EVs are reporting their surrounding information to control centres and training bodies. This not only impact the privacy of surrounding EVs, but also impact the privacy of surrounding pedestrians and infrastructures as well. Since modern EVs are scanning the presence of pedestrians around and for certain functionalities, EVs are also supposed to be connected to the surrounding pedestrians. In this way, if an attacker gets into the collected data, it can easily find out the presence or absence of a particular person/pedestrian at the time of any event. All this reporting and privacy leakage comes under the context of content privacy leakage, which basically means that any information, video, document, picture at a specific interval can cause breach of privacy because of exposure~\cite{intref05}. Therefore, it is important to ensure that the privacy of pedestrians and surrounding infrastructure does not gets leaked at the time of reporting via EVs.



\section{Integration Scenarios of Privacy Preserving Machine Learning in Electric Vehicles}

In the previous sections, we first discuss the motivation that how machine learning is playing a significant role towards the development of modern EVs. Afterwards, we discuss that why privacy preservation is important while developing machine learning models for EVs. Thus, in this section, we highlight the integration scenarios in which PPML needs to be employed in order to efficiently protect privacy of participants in modern EV network.
\subsection{Privacy Preserving Resource Management}
Due to the dynamic nature and heterogeneous structure for EV communication, one of the critical requirement of modern EV networking models is to carry out efficient resource allocation in the network. Resource management covers a wide-range of phenomenon such as handling computing resources, spectrum selection, and transmitting power. Traditional resource allocation models only functions over spectrum resource allocation and does not provide efficient solution to other resource allocation problems. For instance, high mobility EV networks need a rapid allocation response, due to which traditional allocation models are left with short period for optimization result. Thus, in order to carry our efficient resource allocation, modern machine learning models in combination with various enabling technologies are being used~\cite{evinteg01}. For example, integration of machine learning (especially reinforcement learning) models with dynamic EV environment can give efficient outcomes due to their feature of numeric reward maximization. One such example is the integration of Q-learning based power allocation due to crowdsensing in vehicular networks~\cite{evinteg02}. \\
Nevertheless, machine learning aids a lot in carrying out efficient resource management in EVs, however, it also comes with a risk of privacy leakage during this training. For instance, whenever the data of an EV is collected to allocate resource, then various personally identifiable information (PII) associated with this information collection also gets leaked, such as location, identity, etc. Similarly, when the model learns from the data, there is still a possibility that it learns various PIIs, and it can lead to various malicious activities associated with the data. Therefore, it is required to protect privacy at the time of resource allocation. This can be done in both manner, either at the time of data collection or at the time of machine learning based analysis. 

\subsection{Privacy Preserving Routing Decisions}
Traditional routing mechanisms uses local information of EVs (such as relative mobility, location, etc) to perform optimal routing decisions in the network. However, these traditional static optimization models are not good enough in case of modern EVs communicating a huge amount of data~\cite{evinteg10}. Therefore, the trend is shifting towards development of machine learning based routing models, which take in account all possible options before taking a final decision. For instance, machine learning models can learn from multiple features collected via EVs and then can predict an optimal venue on the basis of collected data. One similar work has been carried out by authors in~\cite{evinteg11}, who developed a hierarchical routing mechanism on the basis of Q-learning by taking into account the neighborhood traffic. Similarly, the machine learning based routing models can take into accounts various patterns and associations, which are not possible for traditional routing models. Therefore, machine learning based routing models are being developed and integrated in modern EVs for efficient routing.\\
These machine learning models provide an efficient solution as compared to traditional models, but on the other end, they learn way too much from the data, which can lead to risking of users’ information. For example, in one of the work discussed above, the model learns from neighborhood traffic data. However, if all EV are reporting information about their neighbouring vehicles, then this information can be used to identify the presence of a specific EV on the road at a specific time on a specified location. Similarly, predicting network traffic conditions on one hand is beneficial, but on the other hand it can attract attackers and hackers about a prospective time to attack the network. Therefore, it is significantly critical to preserve privacy of network and routing mechanisms while development of machine learning based routing and networking models. In order to do so, various privacy preserving models, which taken into account the randomness introduction can be used, via which one can still plan the efficient route for a network traffic but can still prevent privacy leakage.

\subsection{Privacy Preserving Intrusion Detection}
Modern EVs carries out its operations with the help of vehicular ad hoc networks also known as VANETs. Nevertheless, VANETs are vulnerable to a large number of attacks, such as eavesdropping, man-in-the middle, sybil attacks, etc~\cite{evinteg12}. Therefore, in order to successfully mitigate these attacks, it is important to implement such mechanisms/detection systems which effectively detect these attacks in a timely manner. To resolve this issue, intrusion detection systems (IDSs) came into discussion, which provides facility to detect the maliciously behaving peers/EVs in the network~\cite{techref10}. A large number of intrusion detection models have been designed which vary on the basis of their functioning, such as statistical anomaly based, collaborative detection, signature based, etc. Similarly, they use various types of machine learning models to predict the outcome for an effective measurement. Though, various types of intrusion detection models are there, but the end goal of all these intrusion detection models is to detect and prevent any adversary or malicious activity in a timely manner.\\
In order to find out the presence or absence of adversaries in an EV network, these models have to collect data from EVs and then they are in a state to carry out processing over the collected data. However, one of the underlying barrier in this data collection is the hesitation of data sharing from EVs because of prospective risk of privacy leakage. For instance, in collaborative intrusion detection, if all nodes share their data in a collaborative manner and some node turns out to be malicious, then it can try to learn private information from the data by eavesdropping in it. Therefore, it is important to carry out integration of a privacy notion in EV data collection or processing at the time of intrusion detection in the network.  

\subsection{Privacy Preserving Crash Avoidance}
EVs are attracting a huge amount of attention because of a number of safety features they are providing. One of the most prominent safety feature is the modern EVs is crash prediction and avoidance. EVs and various control centres take into account the data collected from GPS, pedestrians, traffic signals, and other EVs and predict the occurrence of accident/crash on a specific place~\cite{evinteg03}. Similarly, machine learning models in EVs are also being used to predict unsafe manoeuvres, on the basis of road conditions, number of accidents, time of day, etc. Various machine learning models, such as random forest, principal component analysis, naïve bayes, decision-tree, etc are being used in order to perform efficient predictions~\cite{evinteg04, evinteg05, evinteg06, evinteg07}. This integration provides a subtle number of benefits to EVs and traffic moderation, however, the collection of this data is  not completely secure and can be used for various malicious purposes.\\
The data collected from EVs for the purposes of crash prediction and avoidance comes up with various privacy concerns associate with it. For example, in order to predict a dangerous manoeuvre, the behaviour of EVs needs to be observed for that particular patch of road at a specified time. No doubt, this observation gives fine grained data, which can lead to efficient predictions. But on the other hand, if this data gets in hands of some intruder or malicious node, then it can be misused to predict the presence of a particular EV on a specific place, which can lead to identity privacy leakage for EV users. Similarly, any malicious node can use the collected information to their personal gains, such as spying on a specific EV or finding out the owners of EVs at a particular time when a specific incident occurred. Therefore, in order to provide efficient, private, and secure crash avoidance, it is significantly important to combine the notions of privacy preservation with machine learning models. 

\subsection{Privacy Preserving Trajectory Planning/Route Optimization}
Since EVs are shifting towards autonomous driving, thus, another aspect similar to crash avoidance in EVs is efficient route/trajectory planning. Route planning is not simple as it takes into account a huge amount of data parameters, such as traffic information, obstacles on the road, weather conditions, road conditions, driving comfort, etc~\cite{evinteg13}. All this information is fed to route planning models, which optimize and give the best route from point A to point B. Traditional route planning models/navigators only use built-in maps to plan and give route options, however, modern EVs are equipped with machine/deep learning models, which take into account all above factors alongside the data collected from the onboard units of all EVs, who are on a similar route. As EVs are continuously communicating with each other and RSUs, thus these values are fed into machine learning models for efficient prediction. This type of route planning provides a desirable outcome, as in case if there is any blockage on the road, or there is an undesirable situation, then these EVs can take an alternate route without wasting time on a congested routed.\\
All the values collected from EVs onboard units are super helpful in determining the best and alternate route. Although, the risk of losing the privacy during this collection cannot be undermined at all. For instance, if a fleet of EVs is reported on a narrow road, then an attacker can identify the presence or absence of a particular EV. Similarly, the EV which is using the route planning model has to share its location and destination to machine learning model for the optimized route. Alongside this, if there is any obstacle on the road or a group of people are surrounding a car accident, then the cameras mounted over EVs will be reporting the footage of these people. All these reported and collected values can be misused if they gets into hands of some malicious adversaries, therefore, in order to ensure trust in the EV network, it is significantly important to use privacy preserving techniques at the time of route/trajectory planning. 

\subsection{Privacy Preserving Charging/Discharging Scheduling}
EVs require to charge themselves in order to carry out their operations in a timely manner, and with the advancement of V2I communication, modern EVs are regularly communicating with their nearest infrastructure to schedule their charging period efficiently. For example, an EV owner is going to a shopping center for half hour, so, he/she would like to charge its EV during that period of time. However, its hard to predict that the specified charging spots will be available over the specific time or not. In order to overcome this, the EVs interact with each other and their surrounding infrastructure to schedule them accordingly. Alongside this, it has been proven that EVs also has the capacity that they can discharge themselves in order to provide energy to a specific region at the time of need. This phenomenon of EVs can be used to overcome peak demand and sudden blackouts. For instance, a specific charging discharging schedule can be designed for an EV via which it gets charged at off-peak hours and gets discharged at peak hours. This phenomenon of charging and discharging of EVs is being done with the help of sophisticated machine/deep learning models, which take into account the available data and then compute the optimized outcome for EVs~\cite{evinteg09}.\\
Another important phenomenon that charging/discharging infrastructure take into account while scheduling EVs is the forecasting values of load from the perspective of demand side management~\cite{evinteg08}. For example, if a specific area uses high amount of load in a specified time-slot, then scheduling models will try to move EVs to that specific location for discharging and will try to move to a less demanding location for charging. In order to convince and facilitate EVs, various incentives models are designed. All this work is being done with the help of machine/deep learning models, which take into account all the available data from EVs, charging stations, power grid, etc. and predict and schedule EVs in a way that they meet the desired demand for energy accordingly. \\
No doubt, machine learning models are playing an active role in both charging/discharging and load optimization scenarios for EVs. However, this comes with a cost of prospective privacy leakage in case of collection and training of fine-grained data in traditional machine/deep learning models. A large number of parameters, such as EV location, infrastructure availability, EV availability, etc. are required to compare and compute the efficient schedule for EVs and charging stations. If we analyse these required parameters, it can easily be identified that all these parameters can be termed personally identifiable information parameters. Thus, it will not be convenient for each EV and charging station to share their fine-grained data for the purpose of efficient scheduling. This can only be done if we provide a sense of trust to the corresponding participant that their data will be handled and trained in a private manner. In order to do so, we need to integrate privacy preserving models, both at the time of data collection and at the time of data training.

\subsection{Privacy Preserving EV Energy Bidding/Auction}
Apart from traditional charging/discharging in EVs another well-observed phenomenon that modern EVs are also capable of is trading the energy via the process of online auction. In this process, participants of the auction process submit their bids and asks for an energy commodity, it could be a charging/discharging slot allocation or a replicable battery for plug-in hybrid EVs~\cite{evinteg14}. Once the auction process gets completed, the corresponding buyers and sellers gets the desired commodity. These auctions mechanisms can be pretty complex as well, for instance, in a second price EV auction, the highest bidder wins the auction but pays the second price. Similarly, in a VCG auction, the buyer pays the price equal to the harm it has caused to the network. Therefore, in order to get the most out of these auction mechanisms, research works have identified that machine learning can be integrated at various steps. Firstly, at the time of pre-auction collection of samples, machine learning can be used to estimate the behaviour of participants. After that, machine learning can be used to estimate the revenue and the effectiveness of selling/buying energy from a specific slot/charging station. Finally, after the auction, various statistical analysis can be performed to classify buyers, sellers, and prosumers accordingly. \\
All these models integrating machine learning at various steps of EV auction involve collection of users data, which can be a major cause for the breach of privacy. For example, in the first step, when pre-analysis over data is performed via machine learning. In that case, machine learning models are usually fed with the complete database of all previous auctions, which have records of all participants from the past. Similarly, at the time of auction and bidding analysis, users do not want to reveal their actual bidding/valuation values, but if one gets to analyse the actual results, then bids can be estimated and predicted with exact confidence. Similarly, after auction analysis also involves dealing with users’ private values to predict the future possibilities. Nevertheless, it will not be wrong to say that all these steps can cause privacy leakage accordingly. Therefore, it is critically important to integrate a privacy preserving notion at the time of learning. One way could be to anonymize the datasets before training, while the second direction could be to carry out training in a private manner, such as addition of randomness in the output auction results.

\subsection{Privacy Preserving Autonomous Driving}
Unlike traditional vehicles, autonomous EVs have two major features; namely, the capability of autonomation and cooperation (also known as connectivity)~\cite{evinteg15}. In order to provide this autonomy, these EVs uses latest machine/deep learning models, which are carrying out training over the collected data in a real-time manner. For instance, majority of EVs are always connected with their corresponding control centres and are continuously transmitting the real-time values to them for the purpose of training. The data which is being transmitted mainly depends upon the type and model of EV. Like, if an EV is doing its autonomous driving on the basis of Lidar (light detection and ranging) system, then it is continuously sensing the sensed values. Contrarily, if an EV is taking its decision on the basis of real-time images captured from the camera of EV, then these values are transmitted. Usually, an EV is equipped with more than one type of reporting and controlling mechanisms, such as camera, radar, lidar, ultrasonic sensors, etc. Thus, it will not be wrong to say that an autonomous or even a semi-autonomous vehicle is continuously sending the values from more than one perceiving sensor to a control sensor for training and execution purpose.\\
The data collection from these EVs is extremely important, as it is used to train the models in order to ensure that the decision which EVs are taking are optimized enough in real-time scenarios. However, from the other perspective, if this collected data gets leaked or someone carries out man-in-the-middle attack while collecting this data, then this can cause serious consequences to security and privacy of EVs and their surroundings. For example, if an EV is reporting real-time videos of its surrounding for autonomous driving, then the privacy of all the surrounding EVs and pedestrian is at risk. Similarly, if any adversary gets control of the collected data via data breach, then in this way the malicious people will have complete set of records for any adversarial activity. Therefore, one needs to protect privacy by integrating privacy preserving notion in such scenario. One way could be to carry out decentralized training over EVs instead of a centralized controller. Another way could be to perturb EVs data before transmitting it to control centre, which ensures that even if this data gets leaked or any one gets access to the transmission link, still they will not be able to figure out the exact information.
\subsection{Privacy Preserving Battery Maintenance/Optimization}
Since EVs carry out their operation from the electrically powered battery, which is not an everlasting source and it needs to be replaced after a specified interval of time~\cite{evinteg16}. Nowadays, EV companies are working over development of a long-lasting battery for EV, however, still it will need replacement/checking after a specified time interval. Therefore, a regular condition and other parametric check of an EVs battery is equally important. Nowadays, modern EVs are regularly monitoring the status and the health of their battery and are also reporting these values/parameters to the corresponding manufacturers/control centres as well. Apart from the battery health, the charging stations are not that much abundant in majority of cities across the world in comparison with traditional fossil-fuel stations. Therefore, each EV user has to plan things by keeping this aspect in mind as well. For instance, an EV user has to consider the availability of charging station and battery percentage before planning any trip, as no one wants that the battery of its EV to run out in the middle of road. Nowadays, both of these functions are efficiently being handled by machine learning models. For instance, as EVs are continuously reporting their battery health to manufacturing control centres, thus, a manufacturer can predict with confidence that when battery of an EV is going to die. \\
In order to efficiently monitor and predict both of the above perspective, manufactures are using machine learning algorithms. For instance, health status of battery of an EV is continuously being fed to the model to make efficient predictions. Similarly, the battery status/charging percentage is being fed to such algorithms to predict the presence of a specific amount of power in an area for demand response management, etc. This integration of machine learning with battery related issues of EVs is pretty useful, but it has serious privacy concerns attached with it. The least one is that an EVs’ manufacturer can show targeted advertisement to a customer if they find out that the battery is going to die soon. Moving a step ahead, if this data gets in hands of some adversary, then can identify that when the battery of an EV is going to get down and which charging station it must be targeting. So, the adversaries can easily track EVs on the basis of their battery/charging status. Keeping in view all this discussion, it will not be wrong to say that we need to integrate privacy preserving models with machine learning based battery analysis in EVs. 

\subsection{Summary and Lessons Learnt}
Since the introduction of modern EVs, an unstoppable trend of shifting towards electrified vehicles can be seen in the community because of multiple reasons, such as less-energy cost, reduction in fossil fuels usage, etc. This increase in trend has also attracted attention from both academia and industry, and a large number of industries and researchers are working over development of state of the art protocols for EVs. By examining the works carried out in the development of modern EVs so far, it will not be wrong to say that machine learning and its predecessor similar models have played a great part in enhancing capabilities of EVs. Ranging from basic statistical analysis to automating driving on roads, majority of these tasks are being handled by machine learning models at the backend. However, the data collected for training of these models is not completely harmless as it comes with the risk of privacy leakage, and if not handled properly it can cause serious consequences.\\
The integration of privacy preservation machine learning in EVs is a well-discussed topic, however, a large number of direction still need further exploration. In this section, we summarise some of the critical scenarios, where privacy preservation models needs to be integrated in order to ensure trust in the EV network. One of the most critical scenario that involve data from all sorts of participants in the EV network is resource allocation and management. Traditional machine learning models in EVs take into account various data parameters while allocating resources, but usually the aspect of privacy is missing among them. Apart from resource allocation, another significant aspect from communication network perspective if the routing decisions. For instance, traditional machine learning based routing models in VANETS use a large number of parameters such as location, mobility, destination, trajectory, etc. of an EV before taking routing decisions. However, if all this data get into hands of adversaries, it can cause serious consequences, therefore, privacy is required. Another important aspect that needs serious attention from researchers and industry is the detection of intrusion in EVs via machine learning. Nevertheless, machine learning models work efficiently in detection of intrusion in the network. But in order to predict the presence of an intrusion, we need to feed a hefty amount of personal data into the network, which can cause privacy breach if not handled properly. \\
From the perspective of crash avoidance, researchers are actively working over development of machine/deep learning models, but the perspective of privacy during sensors reporting needs to be analysed further. Similar is the case for trajectory planning, as a large number of sophisticated machine learning models have been designed, however, the aspect of privacy protection while trajectory planning needs a significant attention. Alongside trajectory planning, the direction of charge and discharge scheduling also needs considerable attention so that private schedules can be designed for EV which are concerned about sharing their PIIs. While designing private schedule for charging and discharging of EVs, the aspect of private energy auctions cannot be ignored as they serve as a backbone of charging/discharging phenomenon from incentivization perspective. Apart from this, a plethora of works from the perspective of autonomous driving is available, but the integration of privacy for its data training are not much evident. Therefore, the privacy notion needs to be integrated during the training and collection of data from EVs. Finally, from the direction of EV battery maintenance and replacement, machine learning based predictions are being carried out to estimate the usage and life of battery, but the aspect of private predictions is not considered yet and it needs further attention.


\begin{figure*}[t]
\centering
\scriptsize
\begin{tikzpicture}
\node [block,  text centered, fill=cyan!60, minimum width = 12em,  text width=18em] (a1) {\textbf{Privacy Preserving Machine Learning for Electric Vehicles}};

\node[block, below of=a1, xshift=-18em, yshift=-4.7em, text width=5em](b1){Private Energy Management};
\node[block, below of=a1, xshift=5.5em, yshift=-4.7em, text width=5em](b2){Location-based EV Services Privacy};
\node[block, below of=a1, xshift= 19.25em, yshift=-4.7em, text width=5em](b3){Intrusion \& Misbehavior Detection};
\node[block, below of=a1, xshift= 33em, yshift=-4.7em, text width=5em](b4){Adversial Resilency};


\node[medblock, below of=b1, xshift=-15em,   yshift=-4.7em, text width=4em](b11){Enhancing Energy Demand Prediction};
\node[medblock, below of=b1, xshift=-6.75em, yshift=-4.7em, text width=4em](b12){Incenti- vizing EVs via Blockchain};
\node[medblock, below of=b1, xshift= 4.25em,   yshift=-4.7em, text width=4em](b13){Privacy-Preserving EV Charging};

\node[medblock, below of=b1, xshift= 15.25em,   yshift=-4.7em, text width=4em](b14){Traffic-Sign Recognition};

\node[medblock, below of=b2, xshift=-2.75em, yshift=-4.7em, text width=4em](b21){User Privacy for V2G Network};
\node[medblock, below of=b2, xshift= 2.75em, yshift=-4.7em, text width=4em](b22){Coopera- tive Positioning Error Correction};

\node[medblock, below of=b3, xshift=-5.5em, yshift=-4.7em, text width=4em](b31){Data Falsification Attack};
\node[medblock, below of=b3, xshift= 0em, yshift=-4.7em, text width=4em](b32){Storage Enhancing Collaborative IDS};
\node[medblock, below of=b3, xshift= 5.5em, yshift=-4.7em, text width=4em](b33){Intrusion Detection in Traffic Management System};

\node[medblock, below of=b4, xshift=-2.75em, yshift=-4.7em, text width=4em](b41){Privacy Preservation in VCPS};
\node[medblock, below of=b4, xshift= 2.75em, yshift=-4.7em, text width=4em](b42){Privacy Preservation in IoV};


\node[childblock, below of=b11, xshift=-2.75em, yshift=-7em, text width=4.5em](b111){Enhancing FL + Clustering Based Prediction Error \cite{techref01}, \cite{techref02}};
\node[childblock, below of=b11, xshift= 2.75em, yshift=-7em, text width=4.5em](b112){Enhancing Driving Range Estimation \cite{techref03}};

\node[childblock, below of=b12, xshift= 0em,   yshift=-7em, text width=4.5em](b121){Enhancing State of Charge Estimation Accuracy \cite{techref04}};

\node[childblock, below of=b13, xshift= 5.5em,   yshift=-7em, text width=4.5em](b131){Optimize EV Charging Schedules \cite{techref14}};

\node[childblock, below of=b13, xshift= 0em,   yshift=-7em, text width=4.5em](b132){Enhancing efficiency in EV Charging \cite{techref15}};
\node[childblock, below of=b13, xshift= -5.5em,  yshift=-7em, text width=4.5em](b133){Enhanceing Privacy Protection and Achieving Efficient Scheduling \cite{techref17}};

\node[childblock, below of=b14, xshift= 0em,   yshift=-7em, text width=4.5em](b141){Enhanceing Privacy Protection and Communication Efficiency \cite{techref16}};

\node[childblock, below of=b21, xshift= 0em,   yshift=-7em, text width=4.5em](b211){Enhancing Relative Error w.r.t Privacy Budget and Query Range \cite{techref05}};
\node[childblock, below of=b22, xshift= 0em,   yshift=-7em, text width=4.5em](b221){Enhancing GPS Positioning Prediction Error \cite{techref06}};

\node[childblock, below of=b31, xshift= 0em,   yshift=-7em, text width=4.5em](b311){Enhancing Accuracy of Attack Detection and Communication Cost \cite{techref07}};
\node[childblock, below of=b32, xshift= 0em,   yshift=-7em, text width=4.5em](b321){Enhancing Accuracy of Intrusion Detection with ADMM \cite{techref09}};
\node[childblock, below of=b33, xshift= 0em,   yshift=-7em, text width=4.5em](b331){Accuracy Enhancement for Traffic based Intrusion Detection  \cite{techref10}};

\node[childblock, below of=b41, xshift= 0em,   yshift=-7em, text width=4.5em](b411){Enhancing Accuracy w.r.t Communication Round and Privacy Budget \cite{techref11}};
\node[childblock, below of=b42, xshift= 0em,   yshift=-7em, text width=4.5em](b421){Improved Resilency of IoV from Adversial Attack \cite{techref12}};


\node[MLblock, xshift=-34.0em, yshift=-42em, text width=3.3em](mmm)  {\scriptsize Multi ML Models};
\node[MLblock, xshift=-29.7em, yshift=-42em, text width=3.3em](lr)   {\scriptsize Linear Regression};
\node[MLblock, xshift=-25.4em, yshift=-42em, text width=3.3em](nnr)  {\scriptsize Neural Network Regression};
\node[MLblock, xshift=-21.1em, yshift=-42em, text width=3.3em](FedAG){\scriptsize Fed. Avg. Gaussian};
\node[MLblock, xshift=-16.8em, yshift=-42em, text width=3.3em](gps)  {\scriptsize Gaussian Process Regression};
\node[MLblock, xshift=-12.5em, yshift=-42em, text width=3.3em](brs)  {\scriptsize Bernoulli Random Sampling};
\node[MLblock, xshift=-08.2em, yshift=-42em, text width=3.3em](svm)  {\scriptsize Sparse Vector Technique};
\node[MLblock, xshift=-03.9em, yshift=-42em, text width=3.3em](fl)   {\scriptsize Federa- ted Learning};
\node[MLblock, xshift= 00.4em, yshift=-42em, text width=3.3em](tl)   {\scriptsize Transfer Learning};
\node[MLblock, xshift= 04.7em, yshift=-42em, text width=3.3em](ann)  {\scriptsize Artificial Neural Networks};
\node[MLblock, xshift= 09.0em, yshift=-42em, text width=3.3em](kmc)  {\scriptsize K-means Clustering};
\node[MLblock, xshift= 13.3em, yshift=-42em, text width=3.3em](Lr)   {\scriptsize Logistic Regression};
\node[MLblock, xshift= 17.6em, yshift=-42em, text width=3.3em](nb)   {\scriptsize Naive Bayes};
\node[MLblock, xshift= 21.9em, yshift=-42em, text width=3.3em](ec)   {\scriptsize Ensem- ble Classifier};
\node[MLblock, xshift= 26.2em, yshift=-42em, text width=3.3em](cnn)  {\scriptsize Conven- tional Neural Networks};
\node[MLblock, xshift= 30.5em, yshift=-42em, text width=3.3em](rl)  {\scriptsize Reinfor- cement Learinng};

\path [line] (a1)--($(a1.south)+(0,-0.5)$)-|(b1);
\path [line] (a1)--($(a1.south)+(0,-0.5)$)-|(b2);
\path [line] (a1)--($(a1.south)+(0,-0.5)$)-|(b3);
\path [line] (a1)--($(a1.south)+(0,-0.5)$)-|(b4);
\path [line] (b1)--($(b1.south)+(0,-0.25)$)-|(b11);
\path [line] (b1)--($(b1.south)+(0,-0.25)$)-|(b12);

\path [line] (b1)--($(b1.south)+(0,-0.25)$)-|(b13);
\path [line] (b1)--($(b1.south)+(0,-0.25)$)-|(b14);

\path [line] (b2)--($(b2.south)+(0,-0.25)$)-|(b21);
\path [line] (b2)--($(b2.south)+(0,-0.25)$)-|(b22);
\path [line] (b3)--($(b3.south)+(0,-0.25)$)-|(b31);
\path [line] (b3)--($(b3.south)+(0,-0.25)$)-|(b32);
\path [line] (b3)--($(b3.south)+(0,-0.25)$)-|(b33);
\path [line] (b4)--($(b4.south)+(0,-0.25)$)-|(b41);
\path [line] (b4)--($(b4.south)+(0,-0.25)$)-|(b42);

\path [line] (b11)--($(b11.south)+(0,-0.25)$)-|(b111);
\path [line] (b11)--($(b11.south)+(0,-0.25)$)-|(b112);

\path [line] (b12)--($(b12.south)+(0,-0.25)$)-|(b121);

\path [line] (b13)--($(b13.south)+(0,-0.25)$)-|(b131);
\path [line] (b13)--($(b13.south)+(0,-0.25)$)-|(b132);
\path [line] (b13)--($(b13.south)+(0,-0.25)$)-|(b133);

\path [line] (b14)--($(b14.south)+(0,-0.25)$)-|(b141);

\path [line] (b21)--($(b21.south)+(0,-0.25)$)-|(b211);
\path [line] (b22)--($(b22.south)+(0,-0.25)$)-|(b221);
\path [line] (b31)--($(b31.south)+(0,-0.25)$)-|(b311);
\path [line] (b32)--($(b32.south)+(0,-0.25)$)-|(b321);
\path [line] (b33)--($(b33.south)+(0,-0.25)$)-|(b331);
\path [line] (b41)--($(b41.south)+(0,-0.25)$)-|(b411);
\path [line] (b42)--($(b42.south)+(0,-0.25)$)-|(b421);

\draw [line,blue]  (b111.south)--(mmm.north);
\draw [line,blue]  (b111.south)--(fl.north);
\draw [line,red]   (b112.south)--(lr.north);
\draw [line,red]   (b112.south)--(nnr.north);
\draw [line,red]   (b112.south)--(FedAG.north);

\draw [line,green] (b121.south)--(gps.north);

\draw [line,black] (b131.south)--(fl.north);
\draw [line,black] (b131.south)--(rl.north);
\draw [line,blue]  (b132.south)--(rl.north);
\draw [line,red]   (b133.south)--(rl.north);
\draw [line,green] (b141.south)--(fl.north);

\draw [line,black] (b211.south)--(brs.north);
\draw [line,black] (b211.south)--(svm.north);
\draw [line,blue]  (b221.south)--(fl.north);
\draw [line,blue]  (b221.south)--(tl.north);

\draw [line,red]   (b311.south)--(fl.north);
\draw [line,red]   (b311.south)--(ann.north);
\draw [line,green] (b331.south)--(kmc.north);
\draw [line,green] (b331.south)--(fl.north);
\draw [line,black] (b321.south)--(Lr.north);
\draw [line,black] (b321.south)--(nb.north);
\draw [line,black] (b321.south)--(ec.north);
\draw [line,green] (b331.south)--(svm.north);

\draw [line,blue]  (b411.south)--(fl.north);
\draw [line,blue]  (b411.south)--(cnn.north);
\draw [line,red]   (b421.south)--(fl.north);

\end{tikzpicture}
\caption{\textsc{\\Overview of Privacy Preserving Machine Learning Integration Scenarios in Electric Vehicles}}
\label{fig:currentwork}
\end{figure*}


\begin{table*}[ht]
\begin{center}
 \centering
 \tiny
 \captionsetup{labelsep=space}
 \captionsetup{justification=centering}
\caption{\textsc{\\A Table Summarizing the Integration of Privacy Preservation Protocol in Electric Vehicles Scenarios based on Machine Learning}}
  \label{tab:currentwork}
  {\color{black} \begin{tabular}{|P{1.1cm}|P{1.0cm}|P{0.3cm}|P{2.6cm}|P{1.6cm}|P{1.7cm}|P{1.2cm}|P{1.4cm}|P{1.1cm}|P{1.0cm}|P{0.9cm}|}
 \hline
\rule{0pt}{2ex}
\centering \bfseries Major Domain & \centering \bfseries Sub Category & \centering \bfseries Ref \# & \centering \bfseries Major \newline Contribution & \centering \bfseries Targeted \newline Problem & \centering \bfseries Technical Enhancements & \centering \bfseries Proposed Framework & \centering \bfseries ML Technique & \centering \bfseries Enhanced Parameters & \centering \bfseries Privacy Approach & \bfseries Dataset  \\
\hline
\multirow{4}{*}{}
\rule{0pt}{1ex} \centering \bfseries Private Energy Management & \centering Energy Demand Prediction & \cite{techref01} & Clustering-based EDL approach to minimize the cost of biased prediction and improving the prediction accuracy & Energy demand prediction for EV networks & Enhanced accuracy of EDP and Communication overhead & Clustering Based EDP &\tabitem FL \newline \tabitem Clustering & \tabitem Overhead \newline \tabitem RMSE \newline \tabitem Accurary & Federated Learning & CS dataset of Dundee city  \\
\cline{2-11}

\rule{0pt}{1ex} & \centering Energy Demand Prediction & \cite{techref02} & An economic efficiency framework to maximize the profit margins for all CSs in a vehicular network via FL and Contract Theory & Energy demand prediction and profit maximization for CSs & Enhanced accuracy, learning time, privacy disclosure \& communication overhead & CS-based DFEL & \tabitem Feedforward NN \newline \tabitem FL \newline \tabitem Clustering & \tabitem Utility \newline \tabitem Social Welfare & Federated Learning & CS dataset of Dundee city  \\
\cline{2-11}

\rule{0pt}{1ex} & \centering Driving Range Estimation & \cite{techref03} & Presented the use case of FedAG to carry out the prediction of energy demand of a BEV on given route & EDP and driving range estimation for EVs & Enhanced RMSE and continuous ranked probability score & FedAG & \tabitem LR \newline \tabitem NN Regression \newline \tabitem FedAG \newline \tabitem Clustering & \tabitem Accuracy \newline \tabitem RMSCE & Federated Learning & N/A  \\
\cline{2-11}

\rule{0pt}{1ex} & \centering SoC Estimation & \cite{techref04} & Private and secure energy trading framework and social welfare for IoEV-based demand response & Incentivizing demand response for EVs via blockchain & Enhancing state of charge estimation accuracy & Consortium blockchain-enabled secure energy trading & \tabitem Gaussian Process Regression & \tabitem Utility \newline \tabitem Social Welfare  & Blockchain & N/A  \\
\cline{2-11}

\rule{0pt}{1ex} & \centering Privacy-Preserving EV Charging & \cite{techref14} & FedSAC, a privacy-preserving EV charging control approach using Federated Reinforcement Learning & optimizing EV charging schedules for cost, driver satisfaction, and grid stability while protecting individual charging data privacy & Enhanced convergence effect and generalization ability & Federated EV Charging Control & \tabitem Federated Reinforcement Learning & \tabitem Cumulative Average \newline \tabitem Standard deviation and decline ratio  & Federated Learning & N/A  \\
\cline{2-11}

\rule{0pt}{1ex} & \centering Privacy-Preserving EV Charging & \cite{techref15} & Efficient and private scheduling of EV charging & Cost-effective EV charging with privacy concerns & Scheduling with reinforcement learning and privacy & RL-based EV charging with privacy & \tabitem Reinforcement Learning & \tabitem GA \newline \tabitem Noise function & Adding noise to schedule & N/A  \\
\cline{2-11}

\rule{0pt}{1ex} & \centering Privacy-Preserving EV Charging & \cite{techref17} & Three-level DRL framework for privacy-preserving EV charging scheduling & Balancing privacy of EV arrival/departure times and EVCS energy consumption with efficient charging schedule optimization & Obfuscation of arrival/departure times, revenue maximization for EVCS, operational cost minimization through ESS management & Three-level DRL with Level 1 (privacy-preserving scheduling), Level 2 (multi-agent revenue optimization), Level 3 (ESS-based cost minimization) & \tabitem Reinforcement Learning & \tabitem Privacy parameter $(\varepsilon$) & Differential Privacy, Data Hiding, Obfuscation & N/A  \\
\cline{2-11}

\rule{0pt}{1ex} & \centering Traffic Sign Recognition in autonomous vehicles (AVs) & \cite{techref16} & Privacy-preserving Federated Learning for traffic sign recognition in AVs & Privacy concerns in traditional training methods for AVs requiring raw data sharing & Gradient encryption to protect gradients instead of raw image data & GeFL: System with AVs and server for collaborative training with encrypted gradients & \tabitem Federated Learning & \tabitem Accuracy & Gradient Encryption & N/A  \\
\cline{2-11}
\hline

\multirow{2}{*}{}
\rule{0pt}{1ex} \centering \bfseries Location-based EV Services Privacy & \centering V2G Services Privacy & \cite{techref05} & Presented Quadtree-based algorithm in order to protect the locational privacy parameters of EVs & User privacy for V2G network & Enhanced relative error w.r.t to privacy budget and query range & quadtree-based spatial decomposition algorithm & \tabitem Bernoulli Random Sampling  & \tabitem Relative Error & Differential Privacy & Beijing EVs data \\
\cline{2-11}

\rule{0pt}{1ex} & \centering Cooperative Positioning & \cite{techref06} & A vehicular cooperative position correction system based on federated learning & Cooperative positioning error correction & Enhanced GPS positioning error prediction and convergence speed & FedVCP & \tabitem Federated Learning & \tabitem MAE \& MSE & Federated Learning & Didi Chuxing GAIA Initiative\\
\cline{2-11}
\hline
\multirow{3}{*}{}
\rule{0pt}{1ex} \centering \bfseries Intrusion \& \\ Misbehavior Detection & \centering Data falsification prevention & \cite{techref07} & Used federated machine learning approach to identify the position falsification attack & Detecting data falsification  attack in VANETs & Enhanced precision, recall, accuracy and communication cost in BSM & PPMDS & \tabitem ANN \newline \tabitem FL & \tabitem Precision \newline \tabitem Accuracy & Federated Learning & VeReMi dataset  \\
\cline{2-11}

\rule{0pt}{1ex} & \centering Storage Enhancing Collaborative IDS & \cite{techref09} & Proposed SP-CIDS using ADMM based DML to enhance the storage and computation efficiency of the IDS & Detecting collaborative/ distributed attacks within VANETs & Enhancing accuracy, storage efficiency and scalability of CIDS & SP-CIDS & \tabitem DML \newline \tabitem Logistic Regression \newline \tabitem Niave Bayes \newline \tabitem EC & \tabitem Accuracy & Differential Privacy & NSL-KDD dataset  \\
\cline{2-11}

\rule{0pt}{1ex}& \centering Collaborative IDS & \cite{techref10} & ML based CIDS to facilitate the collaborative exchange of information and sharing of knowledge in VANETs & Collaborative intrusion detection for VANETs & Enhancing ID accuaracy and Security-Privacy trade-off & PML-CIDS & \tabitem SVM \newline \tabitem Logistic Regression & \tabitem Prediction Accuracy & Differential Privacy & NSL-KDD \\
\cline{2-11}
\hline
\multirow{2}{*}{}
\rule{0pt}{1ex} \centering \bfseries Adversarial Resiliency & \centering Layer-wise Privacy Protection & \cite{techref11} & Improved the resiliency of VCPS to adversarial attack in connected vehicles & Privacy-Preservation in VCPS & Enhanced the accuracy w.r.t communication rounds and privacy budget & Adversarial resiliency in vehicular CPS & \tabitem Federated learning & \tabitem Accuracy & Differential Privacy & MINIST data \\
\cline{2-11}

\rule{0pt}{1ex} & \centering Blockchain and FL based Resiliency & \cite{techref12} & A secure, privacy-preserving and verifiable blockchain empowered FL for IoV & Privacy preservation in IoV & Improve resiliency of IoV from adversarial attack & Blockchain empowered FL for IoV  & \tabitem Federated Learning & N/A & Federated Learning & N/A \\
\cline{2-11}

\hline
\end{tabular}}
  \end{center}
\end{table*}


\section{Current works on Privacy Preserving Machine Learning for Electric Vehicles}
EVs have been existent since long, but the developments in modern ICT technologies gave this domain a much-needed boost and researchers started working over development of state of the art protocols for EVs communication and networking~\cite{evtech01, evtech02}. Similarly, this advancement also helped in designing of autonomous vehicles, which are partially or fully being controlled by human intelligence. All this hustle over EVs, helped shape the traditional and conventional vehicle to a modernized fully connected EVs, that we see around us nowadays. One critical technological advancement that played a major role in all this EV advancement is the integration of machine learning models with various scenarios of vehicular networks. This integration paved the path for autonomous operations in EVs, but on the other hand this also comes up with a major issue of privacy leakage~\cite{evtech03}. Various research works also identified and worked over solving this issue in certain scenarios of EVs. From a broader perspective, the works in the field of PPML in EVs can be categorized in four sub-types namely, energy trading, location privacy, intrusion detection, and adversarial resiliency. In this section, we present a detailed summary and technical analysis over all the works in these categories, in order to provide readers a comprehensive overview of the current works carried out so far. An illustrative figure demonstrating the current works in the field of PPML for EVs alongside identifying the proposed or compared machine learning technologies has been given in Fig.~\ref{fig:currentwork}. Similarly, a table discussing the functioning and various technical parameters of the works in the field of PPML in EVs has been given in Table.~\ref{tab:currentwork}.

\subsection{Private Energy Management}
In comparison with traditional fossil-fuel based vehicles, modern EVs perform a lot higher in carbon emission reduction alongside providing an economical way to save energy, which in turn provides a safe and efficient way to carry out everyday commute~\cite{evtech04}. Because of this, we have witnessed a surge in the increase in popularity of modern EVs in major cities around the world. Similarly, these EVs have a capability to store the surplus energy in batteries in order to use this energy at the time of need~\cite{evtech06}. For example, at a time when there is energy deficit in a specific region, EVs can help in overcoming that deficit by discharging themselves in the grid, in this way EVs can also be used to stabilize the energy demand response, which in turn help the distribution smart grid. Therefore, it is important to predict the energy demand and availability in an efficient manner. Nevertheless, a single EV might not make a huge difference, but scientific studies have proven that a large number of EVs cooperating their charging/discharging with a specified energy market in an aggregative manner can make a huge difference to overcome certain unwilling situations, such as blackouts, etc~\cite{evtech05}. Thus, in order to facilitate this notion of energy demand prediction, forecasting, and their interaction among EVs and energy markets, a large number of machine learning based works have been presented and implemented by industry and academia. These works provide an efficient model to predict energy demand, but they have a serious privacy leakage issue associated with them. Thus, it is required to preserve the privacy of modern EVs and its corresponding participants alongside providing efficient energy forecasting and prediction via machine learning models.\\
One such work from the perspective of enhancing energy demand prediction with the help of federated machine learning in conjunction with electric vehicular networks has been carried out by Saputra~\textit{et al.} in~\cite{techref01}. Authors first modified the traditional forecasting models by proposing their own energy demand learning (EDL) model to carry out efficient and accurate predictions at charging stations’ end with the help of data from a specified area. Since, this model had serious privacy and overhead issues, thus, authors proposed another enhanced model for similar task with the help of decentralized federated learning and named it as federated EDL. The major motivation behind integration of decentralized federated learning was that if we train the model in a decentralized manner, then we do not have to collect private information of EVs, which in turn with reduce the risk of privacy leakage and will enhance trust of EVs in the network. Authors further integrated the proposed model with clustering in order to study the effectiveness. Authors carried out comparison with various other machine learning models to provide the effectiveness, and authors claimed that the proposed model improves the energy demand prediction accuracy by 24.63\% alongside reducing the communication overhead of the network by 83.4\%. Similarly, as an extension of their works, authors integrated the notion of contract theory with decentralized federated learning. Authors took into account the multi-principal one-agent optimization problem in the context of smart grid providers and charging stations interconnected with EVs. With the help of decentralized federated learning, authors ensured that the model only train over the required data to ensure that the privacy does not gets leaked while training. \\
Alongside predicating energy demand, another significant parameter that is usually considered is the driving range estimation. Since the charging stations are not available everywhere, even in the major cities, thus, it is important to estimate the driving range of an EV in an efficient manner before planning a driving route. An interesting work targeting the domain of privacy preserving demand range and energy demand prediction for EVs with the help of decentralized federated learning has been presented by authors in~\cite{techref03}. Instead of using the traditional federated averaging (FedAvg), authors worked over using modified FedAvg-Gaussian (FedAG) for efficient energy demand prediction. In order to provide accurate predictions, authors took into account various data parameters such as maps, traffic values, velocity percentile, etc. However, these values falls under the category of PIIs, therefore, authors used decentralized federated learning to restrict the leakage of privacy in the network. From the given evaluation of experiments, it can be visualized that the published technique provides efficient results as compared to traditional models, such as FedAvg. \\
Another work leveraging the use of decentralized to propose a privacy preserving demand response estimation model for internet of EVs (IoEV) has been proposed by Zhou~\textit{et al.} in~\cite{techref04}. Authors worked over development of a distributed framework for EVs in order to enhance privacy alongside providing incentive-compatibility. In order to do so, authors integrated the notion of consortium blockchain in an energy trading framework for EVs. To enhance the incentivization compatibility a bit further, authors worked over integration of contract-theory with EV energy trading to solve for convex optimization problems. In order to effectively estimate the state of charge, authors explored the notion of computational intelligence. By evaluating all these integrations, authors claimed that their proposed framework efficiently ensures the demand response management of a network by providing effective energy trading in a decentralized scenario. Alongside this, it will not be wrong to say that the work proposed by authors can further be enhanced by integrating more state of the art machine/deep learning for accurate prediction and estimation. \\

Electric vehicle (EV) integration into the power grid necessitates optimizing charging schedules for cost, driver satisfaction, and grid stability. However, traditional methods often require centralized data collection, raising privacy concerns. To overcome these challenges, Qian~\textit{et al.} in~\cite{techref14} proposed FedSAC, a novel Federated Reinforcement Learning (FRL) approach that tackles this challenge. FRL allows EVs to collaboratively learn an optimal charging strategy without sharing individual data. Each EV trains a local model on its data, and these models are then combined using FRL. This enables all EVs to learn a globally optimal charging strategy while preserving privacy. FedSAC operates within a simulated environment, treating EV charging control as a decision-making process. The key lies in its two-step approach. First, individual learning empowers each EV to determine an optimal charging strategy using reinforcement learning on its local data.  Second, collaborative learning leverages FRL to aggregate local models from all EVs. This creates a global model that captures collective knowledge, which is then used to refine individual EV models. This iterative process fosters collaboration without compromising data privacy. Simulations demonstrate that FedSAC outperforms existing algorithms in achieving a stable and efficient charging control strategy while preserving privacy. This research holds promise for advancements in privacy-conscious EV charging management, promoting a sustainable and user-friendly electric transportation system. 

Similarly, Hossain~\textit{et al.} in~\cite{techref15} proposed a novel approach for scheduling electric vehicle (EV) charging that prioritizes both efficiency and privacy. The method leverages reinforcement learning (RL) to optimize charging decisions based on time-varying electricity costs while employing a genetic algorithm (GA) to balance exploration and exploitation. To preserve privacy, they proposed method injects noise into the charging schedule, making it difficult to infer actual energy consumption data. This method offers a significant contribution to private energy management for EVs.\\

In another study to improving privacy-preserving training of traffic sign recognition models for autonomous vehicles Parekh ~\textit{et al.} in~\cite{techref16} proposed GeFL. A system leveraging Federated Learning with gradient encryption to address privacy concerns in training traffic sign recognition models for autonomous vehicles. GeFL enables collaborative training on distributed data from autonomous vehicles without sharing raw images. It achieves this by encrypting the gradients of the model, protecting sensitive information while still facilitating model improvement. Their simulation results demonstrate that GeFL achieves high accuracy while significantly reducing data transfer size compared to traditional federated learning approaches. This work offers a promising approach for privacy-preserving training in autonomous vehicles.\\

In another paper by Lee ~\textit{et al.} in~\cite{techref17} proposed a three-level Deep Reinforcement Learning (DRL) framework for privacy-preserving EV charging scheduling in smart EVCS. The framework addresses the challenge of optimizing charging schedules while protecting user privacy regarding arrival/departure times and EVCS energy consumption. It achieves this through a combination of techniques: Level 1 obfuscates arrival/departure times, Level 2 maximizes EVCS revenue using multi-agent DRL, and Level 3 minimizes operational costs through strategic management of the Energy Storage System (ESS). Their results demonstrated the effectiveness of the framework in achieving near-optimal scheduling with strong privacy protection, making it a promising solution for real-world applications.

\subsection{Location-based Services (LBS) Privacy for EVs}
A large number of services in EVs rely of sharing of accurate location and trajectory of EVs to the corresponding party/authority. For example, in order to get estimation of the nearby point of interests (PoI) such as charging stations, restaurants, etc.  an EV has to share its real-time location in the system~\cite{evtech07}. Similarly, various machine learning models are also being used to predict and estimate the factors associated with the relevant PoIs. This data collection indeed is pretty helpful to determine the relevant PoIs, but this location data can indicate the life patterns of EV users, such as religious affiliations, favourite restaurants, preferred shopping centres, etc. In case if an adversary gets their hands on this critical information or an attacker starts eavesdropping this conversation of EVs, they can infer a lot about the users, which can cause serious and harmful consequences. Therefore, one should protect the privacy of EVs whenever they get in contact with any LBS provider. \\
The research works covering the notion of PPML for LBS in EVs are not much abundant, however, in order to provide our readers a brief overview, we discuss two initial works that can pave the pathway for future works in their prospective direction. One such work using the notion of differential privacy with quadtree, Bernoulli random sampling, and sparse vector technique has been proposed by authors in~\cite{techref05}. The work targets the domain of V2G network, where the vehicles are interacting with grid to use grid-based services, such as charging stations, etc. In order to do so, they have to share their private location, which can cause privacy leakage. In order to preserve this, authors integrated random noise addition via differential privacy in the location reporting alongside leveraging the potential of Bernoulli random sampling and quad tree to minimize the relative error. Another work from the perspective of cooperative positioning in internet of vehicles (IoV) has been presented by Kong~\textit{et al.} in~\cite{techref06}. Authors built the motivation for their work by stating that majority of EV and autonomous driving services needs the location permission from corresponding vehicles, which can leak privacy. Similarly, in cooperative positioning schemes, EVs share the positioning data collected from their sensors in order to provide a travel estimation for future vehicles. However, sharing this data on regular basis can cause trajectory leakage, and nobody wants to share their trajectory information with centralized controllers. Therefore, in order to overcome the privacy risks associated with this, authors integrated the notion of decentralized federated learning, which provides EVs with the facility of training the model for cooperative positioning in a decentralized manner. The proposed work ensured the reduction in error rate alongside providing an efficient precision and convergence rate.

\subsection{Privacy Preserving Intrusion \& Misbehaviour Detection}
Operations of EVs are carried out with the help complex functioning environment, which comprises of sensing, communication, and computation components~\cite{evtech08}. Mainly, the sensing part deals with the data collection from EVs surrounding, while network and communication part deals with transmission of this data to concerned authorities/servers. Finally, the computation part is responsible to perform all arithmetic, computational, and logical operations which are required by EVs during communication and decision making process. Since all these steps are interlinked with the data from vehicular network, therefore, it is extremely important to ensure that the data which is being processed and transmitted is legitimate and is not affected by an intrusion. Since, cyber-attacks, including both node-centric and data-centric attacks are pretty common in vehicular networks, thus, it has become an essential practice of EV network controllers to carry out intrusion and misbehaviour detection in the network. Intrusion/anomaly detection is not a completely novel field as it has been there since long, and a large number of works have been carried out so far. However, the use of machine-learning based models to carry out efficient intrusion detection has surpassed all other techniques so far~\cite{evtech09}. Similar is the case with EVs, as the research works have identified that machine learning based intrusion detection in vehicular networks is one of the most effective way to ensure that the data is free from adversarial cyberattacks~\cite{evtech11}.\\
Detection of intrusion in the network is important, but users are usually cautious to share their private data with the regulatory authorities because the data contains PIIs, which can cause privacy leakage. Therefore, it is important to ensure the privacy of users’ data before carrying out analysis over them. One such work using the notion of privacy preserving  decentralized federated learning has been presented by authors in~\cite{techref07}. Authors specifically worked on overcoming data falsification attack by focusing over false positioning data. In order to do so, authors leveraged the benefits of FedAvg in their decentralized model to ensure and enhance accuracy and precision in detection. \\
Apart from using federated learning for privacy protection, certain works also highlighted that we could integrate randomness during learning with the help of differential privacy protection. This is usually done by generating and adding a pseudo random noise at the time of training/collection of data~\cite{evtech10}. In order to evaluate this integration, a very interesting work from the perspective of privacy preserving intrusion detection with the hep of distributed machine learning and differential privacy has been presented by Raja~\textit{et al.} in~\cite{techref09}. The articles first mentioned the need of PPML, alongside discussing the need to mitigate security concerns and attacks in the network. In order to overcome these issues, authors first proposed a distributed machine learning technique which works over alternating direction method of multipliers, also known as ADMM, which enhances the communication and collaboration among V2V interactions. Alongside designing this machine learning model, authors further worked over integration of differential privacy in collaborative IDS (CIDS), and proposed a secure and private CIDS model, which not only provide efficient outcome, but is also resilient to privacy leakage in the network. Another very similar work targeting the domain of distributed and private CIDS has been presented by Zhang and Zhu in~\cite{techref10}. The work not only provides an efficient system to detect intrusion in the network, but also provides a differentially private method to manage traffic and communication in various network scenarios, such as V2V, V2I, I2I communication, etc. Alongside providing a simulation based analysis, authors proposed theoretical evaluations for their proposed PML-CIDS strategy to ensure that their proposed model outperforms other similar scenarios. 

\subsection{Private Adversarial Resiliency}
Modern vehicular networks, which serves as a backbone for successful functioning of interconnected EVs are vulnerable to a number of adversarial attacks. Similarly, with the advancements in modern ICT techniques for EVs, the attackers are also becoming strong and are now trying to develop modern ways to attack a vehicular network in order to get infer into private data of EV users. Similarly, various cybersecurity attacks such as spoofing, reverse engineering, forging, etc. are also becoming very common nowadays. Therefore, apart from preserving specific application/problem in EVs, there is also a need to develop such models which prevent the whole vehicular networks from such attacks alongside preserving the privacy of the network. \\
One such work targeting the notion of attack resiliency in privacy preserving cyber physical systems has been presented by authors in~\cite{techref11}. Authors worked over leveraging the potential of various state of the art technologies, such as federated learning and differential privacy to provide a strong environment under cyberattacks. In order to do so, authors first used FedAvg algorithm to carry out aggregation of model for training and prediction. Afterwards, authors used the notion of layer-wise relevant propagation with differential privacy to preserve the privacy during training at each layer of neurons. In this way, authors ensured that both accuracy and privacy gets maintained for vehicular network. Another very interesting work integrating state of the art notion of blockchain and federated learning with each other to ensure security, privacy, and trust in the IoV network has been presented by authors in~\cite{techref12}. Authors first provided motivation for timely reporting of data in order to take timely decisions, afterwards, authors mentioned the privacy concerns associated with the centralized learning, and in order to eradicate these concerns, authors proposed the use of federated learning in IoV networks. However, authors highlighted that federated learning is also vulnerable to various reverse engineering and poisoning attacks. Thus, in order to make an attack resilient system, authors integrated the notion of immutable blockchain with IoV network. The proposed model ensured that IoV networks provides sufficient amount of security and privacy to the vehicle users. 

\subsection{Summary and Lessons Learnt}
Since the attacks on vehicular networks are increasing rapidly and attackers are trying new ways to overcome the defence mechanisms of vehicular networks. Therefore, nowadays, alongside designing efficient protocols for EVs, it is equally significant to carry out integration of privacy preservation models in these protocols. In order to provide all these integration, various researches have been presented in this direction so far, which we analyse in this section. Firstly, from the viewpoint of privacy preserving energy management, two major scenarios have been targeted from privacy viewpoint. Firstly, certain research works have designed differentially private models for privacy preserving energy demand predictions, and secondly, certain works have highlighted the use of PPML in EV energy trading. \\
Second most prominent scenario from the viewpoint of PPML in EVs is the privacy leakage in location based services. Since location based services are one of the integral part of vehicular networks, as vehicles needs to regularly report their location in order to get the latest updates and in order to get advantages of collaborative facilities. But on the other hand, this reporting led to privacy leakage, which has been regulated by certain research works, firstly from by integrating the randomness via differential privacy and secondly with the help of decentralized federated learning. Alongside location based services, research works also highlighted the need of privacy preserving intrusion detection in the EV network, so that the intruders and misbehaving nodes can easily be identified in a private manner. In order to do so, various research works leveraging the notion of differential privacy and federate learning has been presented in the literature. Finally, from the viewpoint of private adversarial resiliency in EVs, two works have been published so far, which used differential privacy, federated learning, and blockchain as a medium to enhance trust in the network. \\


\section{Open Issues, Challenges, and Future Research Directions}
The interest in EVs is increasing rapidly in both; academia and industry, and a large number of works are being carried out by researchers in order to enhance the capabilities of EVs. Apart from ICTs, another significant integration that paved the path towards this development is integration of machine learning models with vehicular networks. This integration provided EVs with a capability to think like a human in an artificially intelligent manner, which has eased the decision process in EVs. In order to train this machine learning based brain to take autonomous decisions, we need a surplus amount of data from EVs  and its relevant domains/applications. However, if we look into data collection and training perspective, its not fairly easy to collect a huge amount of data from users, as it comes with a number of privacy risks associated with this. To reduce these privacy concerns, various research works have been carried out so far, although still there is plenty of room for future works. In this section, we highlight certain challenges and their corresponding future research work directions for PPML in EVs.

\subsection{PPML Vehicle to Grid (V2G) Network}
Traditional power grid was a one-directional grid, where power grid used to supply generated energy to the affiliated homes and industries. However, with the development of communication technologies, traditional grid has evolved into a modern day bi-directional smart grid, which is efficiently communicating and exchanging energy with its corresponding participants~\cite{evfuture01}. Same is the same with the interactions between smart grid and EVs via V2G network, previously there was just limited interaction. However, nowadays, vehicles and grid are continuously exchanging information related to power/energy flow in the network in order to efficiently manage demand response~\cite{evfuture02}. Alongside this, machine/deep learning models are also being integrated with V2G network to provide effective prediction and forecasting in various scenarios. For example, machine learning models are being used to predict the average charging and discharging units in a V2G network, which in turn help out to carry out efficient load forecasting. Similarly, machine learning based price prediction for V2G auction scenarios and energy markets is being carried out.\\
Nevertheless, these integrations provide help in getting accurate predictions, but the main ingredient behind these predictions is the data of EVs, which can also result in leakage of privacy of EV users and affiliated parties. Till now, very few research works discussed this integration of PPML with V2G networks despite of it being a critical issue. Therefore,  we believe there is a strong need to work over this direction and research works can be carried out from the perspective of integration of state of the art privacy preserving models in EV machine learning. Researchers can study the effect of various privacy preserving techniques such as differential privacy, homomorphic encryption, anonymization, etc. with various machine learning models to figure out and identify the optimal combination for V2G networks. 

\subsection{PPML for VANET Recommender Systems}
The multimedia communication aspect of vehicular networks is enhancing day by day with the emergence of autonomous vehicles, which are capable of providing the feature of self-driving to modern EVs~\cite{evfuture04}. A large number of applications are being developed covering the multimedia aspect of these vehicular networks, and targeted advertisement is one of the most prominent among them~\cite{evfuture03}. These advertisements can for various commercial and non-commercial purposes, such as restaurants, tourism, government messages, etc. In traditional VANETs, advertisements are usually sent to RSUs, and the corresponding RSU shows the ad to the connected EVs, which are within the communication region of that RSU. However, nowadays, state of the art recommender systems are now being used to determine the most suitable ad for each specific EV. For this purpose, these recommender systems use novel machine/deep learning models to allocate the most suitable ad for an EV. In order to do so, the data collected from each EV is put into these recommender systems for training and recommendation purpose. \\
These modernized recommender systems for EVs advertisements give a sophisticated outcome for targeted advertisements, but on the other hand they do also provide a sense of insecurity to EV users. For example, if an EV user is regularly going to a specific type of restaurant, then there is a strong possibility that recommender system will start showing him ads related to the genre of that restaurant. Because of this reason, EV users are usually hesitant to share their location, trajectory, and other similar information to the training servers/authorities. However, if one develop a privacy preserving recommender system for EVs and ensures EV users that their privacy will not get leaked then this issue can be resolved. Therefore, we believe that there is still a plenty of room in this direction and privacy preserving recommendation protocol can be designed for vehicular networks with the help of state of the art notions of privacy, such as differential privacy, etc. 

\subsection{PPML for Cloud-Connected EVs }
With the increase in data-based EV applications, vehicular cloud is becoming an essential part of EV networks, as the data collected from EVs is usually stored over the cloud~\cite{evfuture05}. This is because of the reason that storage hardware over EVs is usually expensive and is not always fully utilized, therefore, in order to maximize utility, researchers moved towards integration of cloud in the vehicular networks. This integration has proved to be fruitful integration, as a large number of applications, such as event sharing, intelligent transportation systems, surveillance, online collaborations, online computing, etc. are now getting benefits out of this integration. Similarly, the integration of machine learning with this cloud-connected EVs scenario has aided EVs further by giving EVs to learn about the behaviour of participating EVs in an efficient manner. Similarly, intrusion/anomaly detection over the cloud data has also become easier with the integration of machine learning protocols.\\
However, in order to get most out of machine learning approaches over cloud-connected EVs, we also need to ensure privacy guarantee over the network. For instance, in case of intrusion detection via machine/deep learning, the database contains data of both type of participants, intruder and normal user. Thus, while identifying the intrusion, the machine learning model also learns attributes of normal user as well, which can lead to identification privacy leakage of individual users in case. Similar is the case with other machine learning based EV-cloud applications because users’ data is directly involved which needs to be protected from any privacy attacks. Therefore, we believe that integrating privacy preserving techniques with cloud-connected EVs during machine learning tasks is the need of future and research works can be carried out over this direction to facilitate EVs. 

\subsection{PPML for Blockchain based EVs }
In the last few years, an exponential interest in decentralized blockchain based systems can be seen due to its immutable and transparent nature. Similarly, in the quest of enhancing trust in the vehicular network, various research works also highlighted the use of blockchain with EVs~\cite{evfuture06}. Since blockchain works over the phenomenon of an immutable ledge, thus, by integrating blockchain with EVs, one can ensure the credibility of data in the system. Similarly, in case of various trading/auction applications of EVs, one can say with confidence that the results are generated in an unbiased manner without any favour, because the trading values are transparent and can be backtracked in a time of need. This indirectly helps in building reputation of each node in the network and can function in highlighting the malicious/misbehaving nodes. In order to facilitate this interaction a bit further, various machine learning models can be designed/evaluated on top of it. For instance, instead of learning from a traditional database, one can learn from blockchain nodes in a decentralized manner with the use of federated learning over blockchain.\\
Alongside getting the most out of this integration, one cannot undermine the risk of privacy leakage during learning and storage over the blockchain. Research works have highlighted that blockchain by its very own nature is a transparent ledger and every node has a copy of the data, thus is has various privacy risks associated with it~\cite{evfuture07}. Therefore, any application/model running over blockchain network also need to ensure that the privacy requirements of the participating users is maintained accordingly. We believe this direction of machine learning over blockchain-based EV network has a lot of potential, but this should be explored by keeping in view the integration of privacy preservation protocols at the time of recording and training of data over and from ledger respectively. 

\subsection{PPML for Cognitive Radio based EV Communicaiton}
In order to carry out vehicular communication, specific spectrum band has been allocated by authorities. E.g., US Federal Communication Commission (FCC) has allocated the spectrum band of 75MHz to VANETS to carry out dedicated short range communication (DSRC)~\cite{evfuture08}. However, various works have highlighted that it is getting congested easily because of the increase in network traffic. One of the solution to overcome this congestion problem is the use of dynamic cognitive radio based spectrum allocation to carry out vehicular communication~\cite{evfuture09}. Cognitive radio works over the phenomenon of allocation of dynamic spectrum to secondary (unlicensed) user when the primary (licensed) user is not using it. Broadly speaking, cognitive radio has four major steps involved named as sensing, analysis, sharing, and mobility, which can usually be carried out by either EVs or RSUs. For instance, if an EV want to communicate via cognitive radio, then first of all it will sense through its environment of figure out the available channels, afterwards, it will carry out analysis of available channel to finalize the optimal one. After that, the EV starts its communication via allocated channel in the sharing phase, and finally, at the time of arrival of PU, the EV move from the selected channel to another one in the mobility phase.\\
These steps are being made more up to date with the help of integration of machine learning with cognitive radio networks. For instance, machine learning models can predict the availability and forecast the parameter at the time of steps involved in cognition cycle. Nevertheless, these steps are facilitated with the help of machine learning, but cognitive radio also suffers with various privacy concerns during cognition cycle~\cite{evfuture10}. Therefore, one cannot directly integrate cognitive radio networks with machine learning based applications of EVs without analysing and overcoming the privacy risks associated with it. This indeed is a good domain, and it has a lot of research potential, but one also need to mitigate the risks of privacy leakage alongside getting the benefits from this fruitful integration.


\section{Conclusion}
The exponential surge in the interest in electric vehicles (EVs) has attracted attention of both academia and industry. Therefore, research works are being carried out to develop optimal models for communication, computation, and other operations of EVs. In order to get most benefits out of EVs, researchers highlighted the use of machine/deep learning models, which effectively use the collected data to make efficient decisions. Nevertheless, it provides tremendous benefits, but it comes with a risk of privacy leakage, which needs to be eradicated in a timely manner. Thus, in this paper, we provide a thorough literature oriented survey from the perspective of privacy preserving machine learning (PPML) in EVs. We first highlight certain basics and preliminaries of privacy preservation, EVs, machine learning, and vehicular ad hoc networks (VANETs). Afterwards, we discuss the motivation of privacy and PPML in EV networks. Then, we highlight certain integration scenarios, where PPML can play an integral role in making the interaction and data storage more secure. After that, we comprehensively discuss the current works carried out in the field of PPML in EVs so far. Finally, we conclude the article by discussing various challenges and future research directions associated with PPML in EVs. 

\bibliographystyle{IEEEtran}


\end{document}